\documentclass[draft]{agujournal2019}
\usepackage{url} 
\usepackage{lineno}
\usepackage[inline]{trackchanges} 
\usepackage{soul}
\usepackage{amsmath}
\usepackage{amssymb}
\usepackage{graphicx}
\usepackage{multirow}
\usepackage{float}
\usepackage{longtable}
\usepackage{booktabs}
\makeatletter
\newcommand\hidejournalhead{\def\@oddhead{}\def\@evenhead{}}   
\makeatother
\AtBeginDocument{\hidejournalhead}
\def\d{{\, \rm d}}

\nolinenumbers
\usepackage{xr}
\draftfalse

\draftfalse

\journalname{}

\begin{document}

%
%


\title{A Stochastic Conceptual Model for the Coupled ENSO and MJO}

%
%




\authors{Yinling Zhang\affil{1}, Charlotte Moser\affil{1}, and Nan Chen\affil{1}}


\affiliation{1}{Department of Mathematics, University of Wisconsin–Madison, Madison, WI 53706, USA}



\correspondingauthor{Charlotte Moser}{crmoser2@wisc.edu}




\begin{keypoints}
    \item The model successfully simulates various El Ni\~no events and captures the strong non-Gaussian statistics across the equatorial Pacific.
    \item The model links intraseasonal, interannual, and decadal variability through dynamic coupling and effective statistical feedback.
    \item The model reproduces multiple MJO-ENSO episodes, including multi-year and extreme events, closely matching observed patterns.
\end{keypoints}

%
%

%
%


\begin{abstract}
Understanding the interactions between the El Ni\~no-Southern Oscillation (ENSO) and the Madden-Julian Oscillation (MJO) is essential to studying climate variabilities and predicting extreme weather events. Here, we develop a stochastic conceptual model for describing the coupled ENSO-MJO phenomenon. The model adopts a three-box representation of the interannual ocean component to characterize the ENSO diversity. For the intraseasonal atmospheric component, a low-order Fourier representation is used to describe the eastward propagation of the MJO. We incorporate a simple decadal process to account for modulations in the background state that influence the predominant types of El Ni\~no events. In addition to dynamical coupling through wind forcing and latent heat, state-dependent noise is introduced to characterize the statistical interactions among these multiscale processes, improving the simulation of extreme events. The model successfully reproduces the observed non-Gaussian statistics of ENSO diversity and MJO spectra while capturing the interactions between wind, MJO, and ENSO.
\end{abstract}

\section*{Plain Language Summary}
El Ni\~no-Southern Oscillation (ENSO) and the Madden-Julian Oscillation (MJO) have been found to interact with each other. In this paper, we develop a stochastic conceptual model to describe such a coupled system. The model is a low-dimensional system with governing equations across multiple time scales. It exploits fundamental dynamical coupling mechanisms and appropriate stochastic ingredients to capture the large-scale dynamics and the statistics of both MJO and ENSO. It succeeds in simulating different El Ni\~no events and tracking MJO propagation and its interactions with ENSO. This model is computationally efficient and statistically accurate, reproducing the observed diversity of El Ni\~no events and their interactions with the MJO. It may help improve understanding of recent variability, providing a foundation for building and improving upon more comprehensive models of the coupled ENSO and MJO.

\section{Introduction}
The El Ni\~no-Southern Oscillation (ENSO) is one of the most significant interannual climate phenomena in the equatorial Pacific, exerting a strong influence on global weather and climate patterns \cite{ropelewski1987global, mcphaden2006enso, latif1998review, neelin1998enso}. ENSO is characterized by quasi-regular cycles in sea surface temperature (SST) across the central and eastern Pacific. ENSO has two phases: El Ni\~no and La Ni\~na, which correspond to positive and negative SST anomalies, respectively. Recent studies have highlighted the diversity and complexity of ENSO \cite{capotondi2015understanding, timmermann2018nino}. ENSO diversity indicates at least two distinct types of El Ni\~no: the eastern Pacific (EP) and central Pacific (CP) events \cite{ashok2007nino, kao2009contrasting, kim2012statistical}. The peak SST anomalies occur in different regions for each type, with EP events centered in the cold tongue region and CP events near the dateline \cite{larkin2005global, yu2007decadal, ashok2007nino, kao2009contrasting, kug2009two}. ENSO complexity further encompasses a range of features, including variations in spatial patterns, peak intensity, and temporal evolution, underscoring the intricate nature of the interannual phenomenon \cite{chen2008nino, jin2008current, barnston2012skill, hu2012analysis, zheng2014asymmetry, fang2015cloud, sohn2016strength, santoso2019dynamics}. Notably, ENSO was identified as a product of tropical air-sea interaction back in the 1960s \cite{bjerknes1969atmospheric}. 

The Madden-Julian Oscillation (MJO) is a prominent intraseasonal climate pattern characterized by large-scale fluctuations in tropical rainfall and atmospheric circulation \cite{zhang2005madden, woolnough2019madden, zhang2020four}. The MJO has been observed to have a significant impact on the ENSO \cite{tang2008mjo, mcphaden2006large, hendon2007seasonal}. The MJO and the convectively-coupled Rossby waves modulate westerly and easterly wind events \cite{puy2016modulation, puy2019influence}. Therefore, they enhance or suppress the development of El Ni\~no and La Ni\~na events by altering wind patterns and convection over the Pacific Ocean \cite{jauregui2024mjo, harrison1997westerly, vecchi2000tropical, tziperman2007quantifying}. Conversely, the SST anomalies, which influence convection and precipitation, affect the amplitude, duration, and spatiotemporal patterns of the MJO \cite{moon2011enso, lee2019enso}. Understanding the interactions between ENSO and MJO facilitates the study of climate variability and improves the forecast of extreme events.

The interaction between the MJO and ENSO is increasingly recognized in climate models, where incorporating this relationship improves both the understanding of model physics and forecast accuracy \cite{mengist2022long, liu2017mjo, ahn2017mjo, hung2013mjo, fernandes2023enso}. Significant progress has been made in recent years toward more realistic simulations of the MJO and ENSO in climate models \cite{khouider2011mjo, ajayamohan2013realistic, guilyardi2020enso, capotondi2020enso}. However, fully capturing the complexities of ENSO, MJO, and their interactions in operational models remains a challenge. Many models in the Coupled Model Intercomparison Project (CMIP), for example, struggle to accurately simulate ENSO diversity, particularly CP events and the strong non-Gaussian characteristics in SST statistics \cite{chen2017enso, dieppois2021enso, capotondi2013enso, atwood2017characterizing}. In particular, most models reproduce eastern-Pacific El Ni\~no events far more readily than central-Pacific ones, and only a few capture both types together. Such model errors often lead to the underestimation of ENSO teleconnections \cite{jiang2021origins, li2019effect, fang2024cmip6}. Similarly, biases in simulating background atmospheric conditions continue to hinder the accurate representation of the MJO in several operational models \cite{kim2017impact, chen2022mjo}.

Conceptual models are valuable tools for breaking down complex phenomena into fundamental components, helping to clarify mechanisms such as feedback loops and interactions between different elements. They also improve the simulation of large-scale features. Insights from conceptual models guide the development of more advanced models, while their accurate simulations and statistics enhance the simulations from more complicated models through multi-model data assimilation and reanalysis \cite{parsons2021multi, bach2023multi, chen2019multi}. Since the 1980s, various low-dimensional conceptual models have been developed to describe the basic oscillatory behavior of traditional ENSO \cite{jin1997equatorial, schopf1988vacillations, wang2017nino, wang2004understanding, mccreary1983model, weisberg1997western, picaut1996mechanism, wang2001unified, chen2023rigorous}. More recently, several new conceptual models based on the recharge oscillator framework have emerged to capture ENSO diversity \cite{chen2022multiscale, geng2020two, fang2024nonlinear}. These models have been used to study the predictability of ENSO's large-scale patterns \cite{fang2023quantifying} and to analyze the statistical response to perturbations in initial conditions and model parameters \cite{andreou2024statistical}. Their statistically accurate outputs have also served as training datasets for machine learning approaches aimed at uncovering critical physical processes \cite{zhang2024physics}. Furthermore, these conceptual models have been used as foundational building blocks for developing simple or intermediate-coupled models (ICMs) \cite{zebiak1987model, geng2022enso, geng2023insights, chen2023simple, thual2016simple}. The MJO, in turn, has been described by simple theoretical models built on several complementary theories, including the skeleton \cite{majda2009skeleton}, moisture-mode \cite{adames2016mjo}, gravity-wave \cite{yang2013triggered}, and trio-interaction \cite{wang2016trio} theories. The skeleton theory, in particular, represents the MJO as a neutrally stable wave arising from the interaction between lower-tropospheric moisture and convective activity, and a stochastic version of it reproduces the intermittency of the observed MJO events. However, these conceptual ENSO models and theoretical MJO models have largely been developed in isolation, and the few existing conceptual models of their coupling still struggle to reproduce realistic ENSO complexity \cite{yang2021enso}. The atmospheric component of the existing ENSO conceptual models is highly parameterized, rather than resolved as an explicit MJO, so that the explicit coupling between the MJO and ENSO has not been adequately addressed in most existing conceptual models. Developing a coupled ENSO-MJO conceptual model would not only bring new insights into natural processes but also facilitate the study of ENSO dynamics, as was highlighted in a recent ENSO recharge oscillator review paper \cite{vialard2024nino}.

In this paper, a stochastic conceptual model is developed to describe the coupled MJO and ENSO phenomena. The ENSO component is represented by a three-box model spanning the equatorial Pacific, capturing large-scale interannual variability across the western Pacific (WP), CP, and EP regions. The model specifically focuses on the CP and EP SSTs, which are critical variables for capturing the diversity and complexity of ENSO events \cite{chen2022multiscale}. For the atmospheric component, a stochastic skeleton model \cite{thual2014stochastic} is utilized as the building block, which has been shown to reproduce many of the observed MJO features. Rather than using a box model, the MJO dynamics in this conceptual model is represented by the leading three Fourier modes from the skeleton model, allowing for a more accurate representation of the eastward propagation of its large-scale features. The coupling between the atmosphere and ocean is modeled such that wind bursts drive ocean currents, while latent heat flux from the ocean influences the strength of convective activity and moisture in the atmosphere. State-dependent noise is incorporated to more effectively capture the feedback mechanisms from the ocean to the atmosphere. The state-dependent noise is crucial for the model to reproduce the observed non-Gaussian statistics. Finally, a decadal variable is introduced into the coupled conceptual model to modulate the background state \cite{capotondi2023mechanisms, power2021decadal}, allowing the model to favor either CP or EP El Ni\~no events with varying frequency over time.

The rest of the paper is organized as follows. Section \ref{Sec:Model} provides a detailed description of the model. Section \ref{Sec:Datasets} outlines the datasets used in this research. The simulation results and corresponding statistical analysis of the coupled model are presented in Section \ref{Sec:Results}. The paper is concluded in Section \ref{Sec:Conclusion}.

\section{The coupled MJO-ENSO model}\label{Sec:Model}

\begin{figure}[h]
\centering
\includegraphics[width=\textwidth]{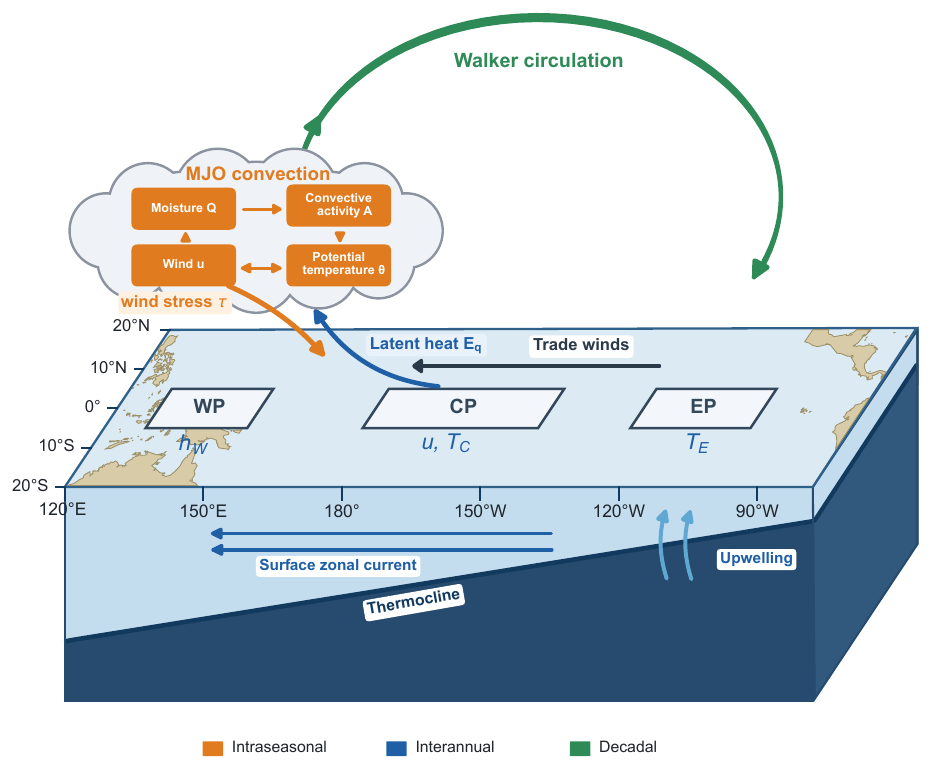}
\caption{A schematic illustration of the stochastic conceptual coupled MJO--ENSO model given in \eqref{Interannual_ocean_model}, \eqref{Intraseasonal_atmosphere_model}, \eqref{Background_A}, \eqref{Total_wind}, and \eqref{Latent_heat_Eq}.}
\label{fig:model_schematic}
\end{figure}

The coupled ENSO-MJO model comprises an interannual ocean component, an intraseasonal atmosphere component, and a decadal variability. ENSO is represented by SST variables in the ocean model, while the MJO is reconstructed from a linear combination of the intraseasonal atmospheric variables. Figure~\ref{fig:model_schematic} summarizes the structure of the model and the coupling pathways that link the intraseasonal, interannual, and decadal time scales. We use the parameter values given in Table \ref{params}.

\subsection{Interannual ocean model}\label{subsec:ocean_model}
The interannual ocean model \eqref{Interannual_ocean_model} is a generalized extension of the classical recharge oscillator model \cite{jin1997equatorial}, applied over the WP, CP and EP to capture both types of ENSO events and their diversity \cite{chen2022multiscale, fang2018three}. This model incorporates the ocean's heat content discharge-recharge processes and zonal advection. In the model, $T_C$ and $T_E$ represent the SST anomalies in the CP and EP, respectively, $U$ denotes the zonal ocean current in the CP, and $h_W$ corresponds to the thermocline depth in the WP. The model reads:
\begin{subequations}\label{Interannual_ocean_model}
\begin{align}
    \frac{\d U}{\d t} &= -r\,U - \frac{\alpha_1 b_0 \mu}{2}(T_C + T_E) + {\beta_u(I)\,\tau},\label{Interannual_ocean_model_U}\\
    \frac{\d h_W}{\d t} &= -r_h\,h_W - \frac{\alpha_2 b_0 \mu}{2}(T_C + T_E) + {\beta_h(I)\,\tau} + C_h,\label{Interannual_ocean_model_h}\\
     \begin{split}
     \frac{\d T_C}{\d t} &= \left(\frac{\gamma b_0 \mu}{2} - c_1(T_C,I,t)\right) T_C + \frac{\gamma b_0 \mu}{2}\,T_E + \gamma_C\,h_W + \rho\,I\,U \\
     &\quad + {\beta_C(I)\,\tau} + \sigma_C\,\dot W_C,
     \end{split}\label{Interannual_ocean_model_T_C}\\
     \frac{\d T_E}{\d t} &= \gamma_E\,h_W + \left(\frac{3 \gamma b_0 \mu}{2} - c_2(T_E,t)\right) T_E - \frac{\delta_{CE}\,\gamma b_0 \mu}{2}\,T_C + C_E + \beta_E(I)\,\tau + \sigma_E\,\dot W_E .\label{Interannual_ocean_model_T_E}
\end{align}
\end{subequations}
Two additional variables couple the interannual ocean to the other time scales. The decadal variable $I$ enters \eqref{Interannual_ocean_model_T_C} through the zonal-advection term $\rho\,I\,U$ and modulates the background state (Section~\ref{subsec:decadal}); the intraseasonal wind $\tau$ enters every equation through the coupling family $\beta_u(I)$, $\beta_h(I)$, $\beta_C(I)$, $\beta_E(I)$, and carries the MJO forcing into the ocean (Section~\ref{subsec:coupling_down}). In addition to the wind-stress forcing, the $T_C$ and $T_E$ equations each contain an additive Gaussian white noise, $\sigma_C\,\dot W_C$ and $\sigma_E\,\dot W_E$, representing fast oceanic forcings that are not transmitted through the wind stress, such as ocean-side eddies and air--sea heat-flux fluctuations. The two damping coefficients $c_1$ and $c_2$ carry the seasonal cycle and the amplitude-dependent nonlinearity, and are specified in Section~\ref{subsec:seasonal}.

\subsection{Intraseasonal atmosphere model}\label{subsec:atmosphere_model}
The intraseasonal atmosphere model generates the intraseasonal wind, a key trigger for El Ni\~no events. Unlike prior studies that used a single process or noise to represent wind in ENSO models \cite{chen2022multiscale, geng2020two, jin2007ensemble}, this model employs simple atmospheric equations to capture more detailed dynamics, including the MJO. The governing equations are based on the stochastic skeleton model for the MJO \cite{thual2014stochastic}. However, only the modes up to the third zonal wavenumber are retained, simplifying the model while still capturing large-scale MJO features \cite{majda2009skeleton, majda2011nonlinear, ling2017challenges}. The governing equations, written in terms of Fourier modes, are as follows,
\begin{subequations}\label{Intraseasonal_atmosphere_model}
 \begin{align}
     \frac{\d\hat{K}_k}{\d t} &= -d_k \hat{K}_k -i k \hat{K}_k -  \overline{H} \hat{A}_k/2, \label{Intraseasonal_atmosphere_model_K}\\
     \frac{\d\hat{R}_k}{\d t} &= -d_k \hat{R}_k + i k \hat{R}_k/3 - \overline{H} \hat{A}_k/3, \label{Intraseasonal_atmosphere_model_R}\\
    \frac{\d\hat{Q}_k}{\d t} &= - d_k \hat{Q}_k - i k  \overline{Q} (\hat{K}_k - \hat{R}_k/3)+\overline{H} \hat{A}_k(\overline{Q}/6 - 1) + \sigma_{\hat{Q}_k}(E_q) \dot{W}_{\hat{Q}_k}, \label{Intraseasonal_atmosphere_model_Q}\\
    \frac{\d\hat{A}_k}{\d t} &=  -\lambda_A \hat{A}_k + \sum_j\Gamma \hat{Q}_{j} (\hat{A}_{k-j} + \bar{A}_{k-j} )+ \sigma_{\hat{A}_k}(Q,A) \dot{W}_{\hat{A}_k},\label{Intraseasonal_atmosphere_model_A}
 \end{align}
 \end{subequations}
where $k=0,\pm 1, \pm2$ and $\pm3$. The four model variables (indicated with hats) on the left-hand side of \eqref{Intraseasonal_atmosphere_model} are the Fourier coefficients of the Kelvin wave ($K$), Rossby wave ($R$), moisture ($Q$), and convective activity ($A$), which are all anomalies. Note that $\bar{A}_k$ represents the Fourier coefficients of $\bar{A}$, which is the background convective activity computed by
\begin{equation}\label{Background_A}
H\bar{A} = E_q + S^q-\bar{Q}S^{\theta}/(1-\bar{Q}),
\end{equation}
where $S^\theta$ and $S^q$ are the external source of cooling and moistening, respectively. Both $S^\theta$ and $S^q$ are pre-determined, peaking in the warm pool region. In \eqref{Background_A}, $E_q$ is the latent heat that is affected by the SST and provides the feedback from the ocean to the atmosphere (Section~\ref{subsec:coupling_up}). The other two factors $\overline{H}$ and $\overline{Q}$ are also constants \cite{thual2014stochastic}. A linear combination of $\hat{K}_k$, $\hat{R}_k$, $\hat{Q}_k$, and $\hat{A}_k$ can also be utilized to reconstruct the wind and the MJO, where a temporal filter within the band 30 to 90 days is further carried out for reconstructing the MJO. Note that this temporal filter is applied only to the reconstructed MJO signal used for visualization and diagnostics. It is not applied to the wind $\tau$ or the underlying atmospheric variables, which retain their full spectral content. Please refer to Appendix~\ref{sec:MJO_reconstruct} for further details.

The skeleton model for the MJO \cite{majda2009skeleton} captures several key observational features: (i) the eastward propagation speed of 5 m/s, (ii) a dispersion rate of $\d\omega/\d k \approx 0$, and (iii) a horizontal quadrupole structure. The stochastic version \cite{thual2014stochastic} further simulates (iv) the intermittent generation of MJO events and (v) the organization of MJO events into wave trains that exhibit growth and dissipation. These properties are retained in the low-order representation provided in \eqref{Intraseasonal_atmosphere_model}. Unlike the box model used for the ENSO component in \eqref{Interannual_ocean_model}, the Fourier representation in \eqref{Intraseasonal_atmosphere_model} enables a more accurate depiction of the eastward propagation of the MJO's large-scale features. Since the modes $k=-1,-2$, and $-3$ are complex conjugates of those $k=1,2$, and $3$, only 16 equations (four for each wavenumber, together with the mode $k=0$) are needed for characterizing the intraseasonal atmospheric component.

The original stochastic skeleton model employed a stochastic birth-death process \cite{gardiner2009stochastic, thual2014stochastic} to capture the irregularity of MJO events. It has been demonstrated that, in the limit of infinitesimal jumps, a continuous stochastic differential equation driven by white noise $\dot{W}_A$ with a state-dependent coefficient can be derived \cite{chen2016filtering}. This formulation facilitates the spectral decomposition of the model and is the one utilized here. Although the system \eqref{Intraseasonal_atmosphere_model} is expressed in Fourier space, the state-dependent noises are easier to interpret when the system is presented in physical space. In physical space, the state-dependent noises in the equations for convective activity and moisture take the following form:
\begin{subequations}\label{Noises_State_Dependent}
\begin{align}
\sigma_A &= \sqrt{\nu|Q|(A+\bar{A})},\label{Noises_State_Dependent_A}\\
\sigma_Q &= \max\left(\tilde{\sigma}_Q \left(1-e^{-c_qE_q}\right), {\sigma_{Q}^{\min}}\right),\label{Noises_State_Dependent_Q}
\end{align}
\end{subequations}
where $\nu$, $\tilde{\sigma}_Q$, $c_q$, and $\sigma_Q^{\min}$ are all positive constants. The noise coefficient $\sigma_A$ in \eqref{Noises_State_Dependent_A}, derived from the limit of infinitesimal jumps in the original stochastic skeleton model \cite{chen2016filtering}, indicates that the strength of the convective activity is dependent on the moisture level. This relationship is consistent with the deterministic part of the model, $\Gamma Q (A+\bar{A})$, where moisture influences the tendency of the convective activity. The moisture noise $\sigma_Q$ links the ocean and the atmosphere via the latent heat variable $E_q$ (Section~\ref{subsec:coupling_up}); a minimum noise in the moisture process, denoted as $\sigma_Q^{\min}$, is prescribed to ensure that the noise coefficient remains non-negative. Since only the leading three Fourier modes are retained, the stochastic noises not only capture the explainable physics but also compensate for contributions from smaller scales.

\subsection{Coupling from the intraseasonal to the interannual dynamics}\label{subsec:coupling_down}
The intraseasonal atmosphere drives the interannual ocean through the MJO wind stress. The wind $\tau$ appearing in \eqref{Interannual_ocean_model} is the intraseasonal wind reconstructed from the leading Kelvin and Rossby modes of the skeleton, averaged over the western Pacific where wind activity is most pronounced:
\begin{equation}\label{Total_wind}
\tau = u_W .
\end{equation}
This wind forcing is incorporated into each of the interannual ocean model equations through the coupling family $\beta_u(I)$, $\beta_h(I)$, $\beta_C(I)$, $\beta_E(I)$, which specify how the intraseasonal wind projects onto the components of the interannual ocean state vector. Specifically, the projection onto $U$ accounts for the wind-driven adjustment of zonal currents, the projection onto $h_W$ corresponds to the wind-driven modulation of thermocline depth, and the projections onto the sea surface temperature variables represent the surface temperature response associated with the coupled ocean-atmosphere adjustment in the equatorial Pacific. The finite time the ocean takes to adjust to the wind is not imposed through the forcing, but emerges through the model dynamics: the wind acts directly on the zonal current, the western Pacific thermocline, and the SST, and the delayed SST response arises through the evolution of the coupled ENSO dynamics and the recharge-discharge evolution. The dependence of the coupling coefficients $\beta(I)$ on the decadal variable is described in Section~\ref{subsec:decadal} and \ref{sec:params}.

\subsection{Coupling from the interannual to the intraseasonal dynamics}\label{subsec:coupling_up}
The interannual ocean feeds back on the intraseasonal atmosphere through the latent heat. The latent heat $E_q$ in \eqref{Background_A} is approximated by a function proportional to the SST in the CP, serving as a bulk representation,
\begin{equation}\label{Latent_heat_Eq}
E_q = \alpha_q T_C .
\end{equation}
As the SST increases, it is expected that convection and moisture will also increase, which modulates the atmosphere. This occurs because a warmer central Pacific enhances surface evaporation and the latent heat flux into the atmosphere, supplying additional moisture and energy for deep convection \cite{xie2020new}. Since the mean deep convection is anchored over the warm pool, it is the central Pacific SST, rather than the colder eastern Pacific where deep convection is suppressed, that determines how far this convective region extends eastward \cite{picaut1996mechanism}, so the latent heat is tied to $T_C$. This feedback operates in two ways. First, $E_q$ sets the background convective activity $\bar{A}$ in \eqref{Background_A}, which appears in the deterministic forcing $\Gamma Q(A+\bar{A})$ of the convective-activity equation \eqref{Intraseasonal_atmosphere_model_A}, so that a warmer central Pacific raises $\bar{A}$ and enhances the convective activity. Second, the same latent heat enters the state-dependent noises, where the convective-activity noise $\sigma_A$ in \eqref{Noises_State_Dependent_A} grows with $\bar{A}$ and the moisture noise $\sigma_Q$ in \eqref{Noises_State_Dependent_Q} grows with $E_q$, so that a higher central Pacific SST strengthens the statistical coupling between the atmosphere and ocean and favors westerly wind bursts ahead of warm events. Together with the wind forcing of Section~\ref{subsec:coupling_down}, these two pathways close the two-way MJO--ENSO coupling loop.

\subsection{Decadal modulation and the triggering of ENSO diversity}\label{subsec:decadal}\label{subsec:triggering}
The decadal variability in the conceptual model is primarily used to modulate the background state over the equatorial Pacific \cite{capotondi2023mechanisms, power2021decadal}, favoring the generation of more EP or CP events over specific time periods. Developing a comprehensive model describing the decadal variation is not the main focus. To this end, a simple linear stochastic process is employed as the governing equation for the decadal variable $I$,
\begin{equation}\label{Decadal_variability}
     \frac{\d I}{\d t} = -\lambda(I-\bar{I}) + \sigma_I(I) \dot W_I,
\end{equation}
where the damping coefficient $\lambda$ is chosen such that $I$ varies in the decadal timescale and $\bar{I}$ is the mean state.
It is important to note that the trade winds in the lower level of the Walker circulation on decadal timescales are predominantly easterly. This implies that the sign of $I$ remains constant over time, resulting in a non-Gaussian distribution of $I$. This feature can be easily integrated into the linear stochastic process of $I$ by using a state-dependent noise coefficient \cite{averina1988numerical, yang2021enso}. Given the limited observational data and the principle of deriving the least biased maximum entropy solution for a distribution \cite{chen2023stochastic}, a uniform distribution is adopted for $I$ in this work. Here, a larger value of $I$ corresponds to stronger easterly trade winds. As the strength of the Walker circulation background, $I$ modulates the coupled system in two ways. It sets the zonal-advection feedback $\rho I U$ in \eqref{Interannual_ocean_model_T_C} that selects the EP or CP regime, and it sets the wind-coupling coefficients $\beta(I)$ in \eqref{Interannual_ocean_model} that determine how strongly the intraseasonal MJO wind forces the ocean. When the Walker circulation is weak (small $I$), the MJO wind exerts a stronger influence on ENSO, whereas when it is strong (large $I$), zonal advection dominates and this influence weakens (see \ref{sec:additional_analysis}). The decadal variable therefore links the intraseasonal, interannual, and decadal scales.

It is worth explaining the model mechanisms for triggering CP and EP El Ni\~no events. In the absence of the stochastic wind forcing, the model is a deterministic and nearly linear system if the decadal variable $I$ is held constant. Note that there is nonlinearity in the damping terms of \eqref{Interannual_ocean_model_T_C} and \eqref{Interannual_ocean_model_T_E}, but they are weak and do not play a significant role when examining the function of the strong nonlinear advection term $\rho\,I\,U$. The model exhibits different oscillatory patterns depending on the fixed value of $I$ \cite{fang2018three}. The value of $I$ indicates the strength of the SST gradient, which directly modulates the efficiency of zonal advection. Since the zonal ocean current is weak in the EP, zonal advection is only significant to ENSO in the CP region. Thus, the nonlinear zonal advection term provides the mechanism to warm the CP without significantly warming the EP. When $I$ is small, corresponding to weak advection, the model reduces to the classic recharge oscillator paradigm and generates the EP El Ni\~no cycle. In contrast, when $I$ is large, the strengthened SST gradient enhances the role of advection, leading to a dominant CP El Ni\~no oscillatory pattern. The time-varying $I$, combined with the stochastic forcing, allows for the occurrence of different El Ni\~no events over time. Depending on the value of $I$, one type of event becomes more likely during a given period, thereby reproducing the observed alternation between EP- and CP-dominant phases \cite{yu2013identifying}. A more thorough exploration of the decadal variability is presented in \ref{sec:additional_analysis}.

\subsection{Seasonal phase locking}\label{subsec:seasonal}
The influence of seasonality has been incorporated into the two damping coefficients, $c_1$ and $c_2$, to more accurately capture the seasonal phase-locking behavior of ENSO, as shown in equations \ref{c1} and \ref{c2}. This behavior is characterized by the tendency of ENSO events to peak during boreal winter \cite{tziperman1997mechanisms, stein2014enso, fang2021effect}. The damping parameter $c_1$ is also a quadratic function of $T_C$ which allows the model to produce the negatively skewed PDF for CP SST anomalies, and its floor depends linearly on the decadal variable $I$ to favor CP El~Ni\~no events under a stronger Walker circulation. Similarly, the quadratic-in-$T_E$ term in $c_2$ makes the effective damping of $T_E$ cubic. From the viewpoint of ENSO as a nonlinear oscillator, this produces amplitude-dependent damping that strengthens as the SST anomaly grows and thereby limits the amplitude of large excursions of either sign \cite{jin1997equatorial, an2004nonlinearity}.
\begin{subequations}
    \begin{align}
        \label{c1}c_1(T_C, I, t) &= 0.65\left[16\left(T_C + 0.05\right)^2 + 1.45 - 0.15\,I\right]\left[1 + 0.45\sin\theta\right],\\
        \label{c2}c_2(T_E, t) &= 0.5\left[12\,T_E^{\,2} + 1.60\right]\left[1 + 0.32\sin\!\left(\theta + \frac{2\pi\,\phi_2}{12}\right) + 0.32\sin\!\left(2\theta + \frac{2\pi\,\phi_2}{12}\right)\right],
    \end{align}
\end{subequations}
with
\begin{equation}\label{c2_phase}
\theta = \frac{2\pi}{6}\!\left(t - \frac{\phi_1}{2}\right),
\end{equation}
where $\phi_1 = 0.5$~months and $\phi_2 = 1.33$~months. This seasonal damping represents the seasonal migration of the Intertropical Convergence Zone, which modulates the upwelling and horizontal advection and thereby the evolution of the SST \cite{chen2022multiscale}. The modulation of $c_1$ uses a single annual harmonic, whereas that of $c_2$ adds a second harmonic. Following previous work, the additional harmonic is included to better represent the observed seasonal modulation of variability in the eastern Pacific \cite{stein2014enso, chen2022multiscale}. The amplitudes and the phase lags $\phi_1$ and $\phi_2$ were chosen so that the simulated seasonal cycle of SST variance matches the observations (Figure \ref{fig:Statistics_ENSO}, Panel (d)).
\section{Observational data and diagnostics}\label{Sec:Datasets}

Observational data is utilized to validate the simulated time series and statistics from the model.

\subsection{Observational data sets}

The monthly SST data used in this study is sourced from the GODAS dataset \cite{behringer2004evaluation}. The SST anomalies are calculated by subtracting the monthly mean climatology over the entire analysis period. The Ni\~no 4 and Ni\~no 3 indices represent the average SST anomalies over the zonal regions $160^\circ$E–$150^\circ$W and $150^\circ$W–$90^\circ$W, respectively, with a meridional average between $5^\circ$S and $5^\circ$N. The reanalysis period spans from 1982 to 2019, corresponding to the satellite era when observations are more reliable. This period is used to study ENSO-MJO spatiotemporal patterns. For statistical comparisons, a longer SST dataset (1950–2019) is used, obtained from the Extended Reconstructed Sea Surface Temperature version 5 \cite{huang2017extended}.

To represent convective activity ($a$), we use the National Oceanic and Atmospheric Administration's (NOAA) interpolated outgoing longwave radiation (OLR) dataset \cite{liebmann1996description}. While several proxies for convective activity exist, OLR is chosen here as an initial, straightforward option for analysis. For other variables such as zonal wind, geopotential height, and specific humidity, we rely on the National Centers for Environmental Prediction-National Center for Atmospheric Research (NCEP-NCAR) reanalysis data \cite{kalnay1996ncep}. Both datasets offer a horizontal spatial resolution of 2.5$^\circ\times$2.5$^\circ$ and are available at daily temporal intervals. The period used for this study spans from 1982 to 2019 \cite{stechmann2015identifying}. The processing of these atmospheric fields into the intraseasonal model variables follows the procedure detailed in Appendix~\ref{sec:data_processing}.

\subsection{Definitions of ENSO events}

The classification of different El Ni\~no and La Ni\~na events, which helps examine ENSO variability, is based on the criteria outlined in \cite{kug2009two, wang2019historical}. These classifications rely on the average SST anomalies during boreal winter (December–January–February; DJF). An event is classified as an EP El Ni\~no if the EP is warmer than the CP, and the EP SST anomaly exceeds $0.5^\circ$C. An extreme El Ni\~no occurs when the maximum EP SST anomaly between April and the following March exceeds $2.5^\circ$C. A CP El Ni\~no is identified when the CP is warmer than the EP, with a CP SST anomaly above $0.5^\circ$C. Lastly, a La Ni\~na event is defined when the SST anomaly in either the CP or EP falls below $-0.5^\circ$C.

\subsection{Definitions of MJO signals}

The MJO signal is reconstructed from the intraseasonal atmospheric state variables introduced in Section~\ref{Sec:Model}. Because the atmospheric component is already expressed through the Fourier coefficients of the Kelvin wave, Rossby wave, moisture, and convective activity, $\hat{K}_k$, $\hat{R}_k$, $\hat{Q}_k$, and $\hat{A}_k$, the MJO signal is obtained by projecting these variables onto the MJO mode of the skeleton model. For each zonal wavenumber $1\le|k|\le 3$, let $\mathbf{X}_k(t)=(\hat{K}_k,\hat{R}_k,\hat{Q}_k,\hat{A}_k)^{\mathtt{T}}$ denote the atmospheric state vector, and let $\hat{\mathbf{e}}_{\mathrm{MJO}}(k)$ be the MJO eigenvector obtained from the linear analysis of the skeleton model \eqref{Intraseasonal_atmosphere_model}, which quantifies the contribution of each characteristic variable to the MJO mode. The MJO signal in Fourier space is then the projection
\begin{equation}\label{MJO_projection}
\widehat{\mathrm{MJO}}(k,t) = \hat{\mathbf{e}}_{\mathrm{MJO}}(k)^{\dagger}\,\mathbf{X}_k(t),
\end{equation}
where $\dagger$ denotes the conjugate transpose. An inverse Fourier transform of $\widehat{\mathrm{MJO}}(k,t)$ over the retained wavenumbers $1\le|k|\le3$ yields the physical-space signal $\mathrm{MJO}(x,t)$, to which a temporal band-pass filter is finally applied to retain only the intraseasonal band of 30 to 90 days. The eigenvalues, the eigenvectors $\hat{\mathbf{e}}_{\mathrm{MJO}}(k)$, and the detailed derivation are provided in Appendix~\ref{sec:MJO_reconstruct}.
\subsection{Spatiotemporal reconstruction}\label{subsec:reconstruction}
In the conceptual model developed here, ENSO is characterized by two SST indices, $T_E$ and $T_C$, as described in \eqref{Interannual_ocean_model}. Similarly, the MJO is represented by the leading three Fourier modes in \eqref{Intraseasonal_atmosphere_model}. Although both phenomena are represented in a low-dimensional form using time series, a spatiotemporal reconstruction based on these time series enables us to visualize the spatiotemporal patterns and spatial propagation of the observed features. The details and validation of the spatiotemporal reconstructions can be found in Appendix~\ref{sec:spatial_reconstruct}.
\section{Model simulation results}\label{Sec:Results}

\subsection{Coupled simulation}

\begin{figure}[h]
\hspace*{-0.3cm}\includegraphics[width=1\textwidth]{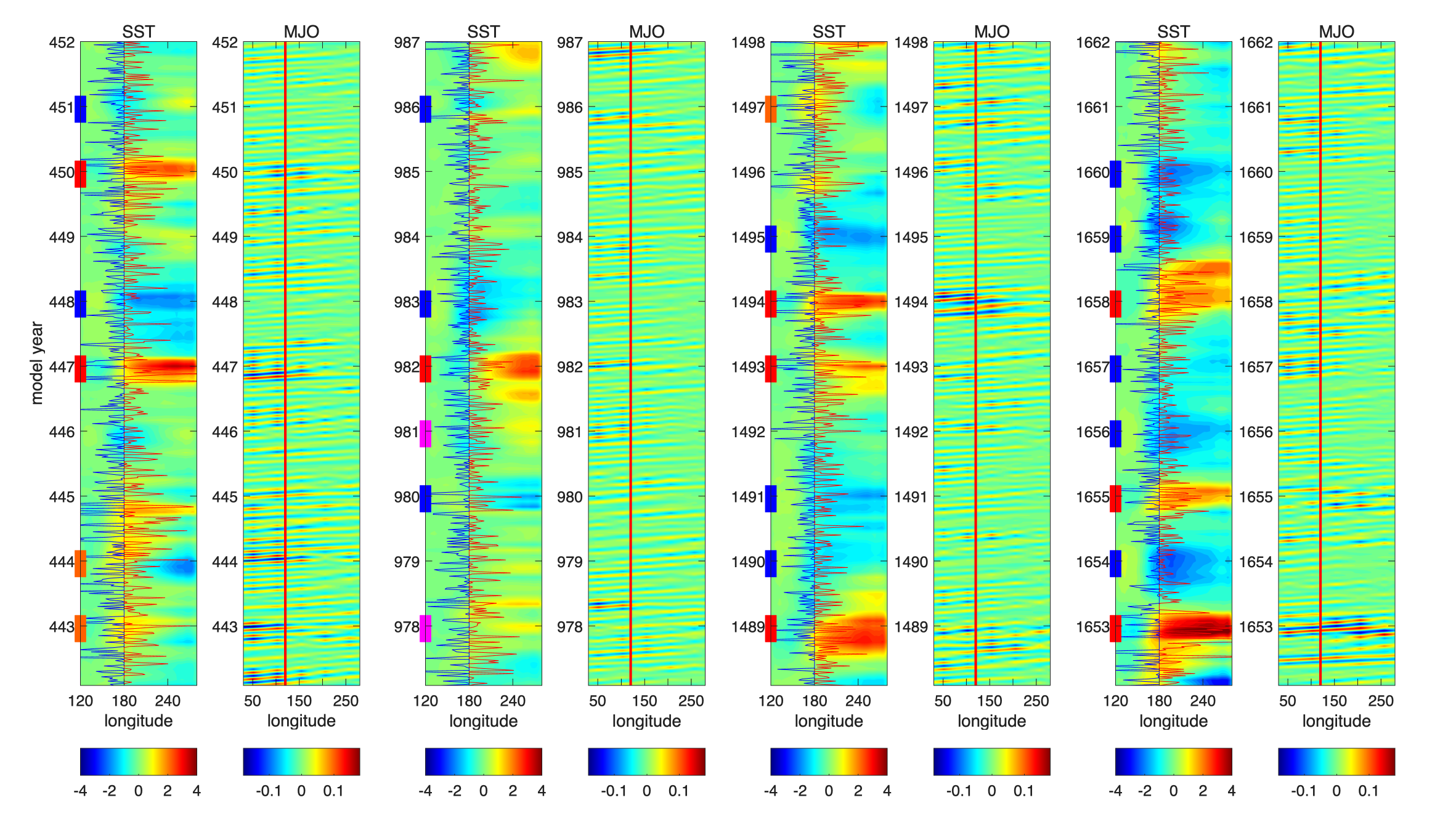}
\caption{Hovmoller diagrams of the spatially reconstructed SST ($^\circ$C) and MJO (unit-less) from the conceptual model. The MJO signal is represented by a linear combination of $K, R, Q,$ and $A$ (see Appendix~\ref{sec:MJO_reconstruct}). The x-axis of the SST Hovmoller diagram covers the equatorial Pacific, while the MJO diagram also includes the Indian Ocean. The red vertical line in the MJO panels marks the Western Pacific (WP) boundary at 120$^\circ$E. In the SST panels, the averaged atmospheric wind over the WP is overlaid on the SST, with red and blue indicating westerly and easterly intraseasonal wind anomalies, respectively. The rectangles along the y-axis indicate different ENSO events occurring during boreal winter: strong EP El Ni\~no (red), moderate EP El Ni\~no (purple), CP El Ni\~no (orange), and La Ni\~na (blue).}\label{fig:Hovmoller_ENSO_MJO}
\end{figure}

Figure \ref{fig:Hovmoller_ENSO_MJO} shows the Hovmoller diagrams of the spatially reconstructed SST and MJO from the conceptual model. These diagrams illustrate the diversity and complexity of ENSO, as well as the coupled relationship between ENSO and the MJO. A similar figure based on observational data can be found in Appendix~\ref{sec:obs_hov}.

First, the model simulates different types of ENSO events with frequencies similar to observations (see also Panel (f) in Figure \ref{fig:Statistics_ENSO}). A few examples of each type, with corresponding years of observed events in parentheses, are as follows: (i) moderate EP El Ni\~no (1983): $t=$978; (ii) extreme EP El Ni\~no (1998): $t=$447; (iii) delayed super El Ni\~no (2014-2015): $t=$1652-1653; (iv) multi-year EP El Ni\~no (1987-1988): $t=$981-982; (v) CP El Ni\~no (2010): $t=$1497; (vi) multi-year CP El Ni\~no (1993-1994): $t=$443-444; (vii) La Ni\~na (2011): $t=$1495; (viii) multi-year La Ni\~na (1999-2000): $t=$1490-1491. Additional examples can be found in Figure \ref{fig:SI_ENSO_Time_Series}.

Second, the model successfully reproduces strong interactions between the MJO and ENSO. Typically, intense MJO events are observed either preceding or during the El Ni\~no phase, for both EP and CP El Ni\~no events. The MJO signals extend further towards the EP, particularly during strong EP El Ni\~no events. However, some MJO events are observed during La Ni\~na phases (e.g., at $t=$1657) or are not associated with ENSO events at all (e.g., at $t=$446). These findings are consistent with observations, as detailed in Appendix~\ref{sec:obs_hov}.

These interactions operate through the two coupling pathways of the model. Ahead of El Ni\~no onset, enhanced MJO activity is accompanied by westerly intraseasonal wind anomalies over the western Pacific (overlaid on the SST panels) that force the ocean and help trigger the warming, most clearly for the extreme EP events. Conversely, subdued MJO activity leaves the intraseasonal wind weak and favors La Ni\~na. In turn, the warm SST during El Ni\~no strengthens the latent heat and drives the eastward extension of the MJO signal toward the eastern Pacific. Because the intraseasonal wind acts as a stochastic and intermittent trigger rather than a deterministic forcing, the relationship between the MJO and ENSO is not one-to-one, and some strong MJO events fall during La Ni\~na or outside any ENSO event.

\subsection{Statistics}

\begin{figure}[h]
\hspace*{-0cm}\includegraphics[width=1\textwidth]{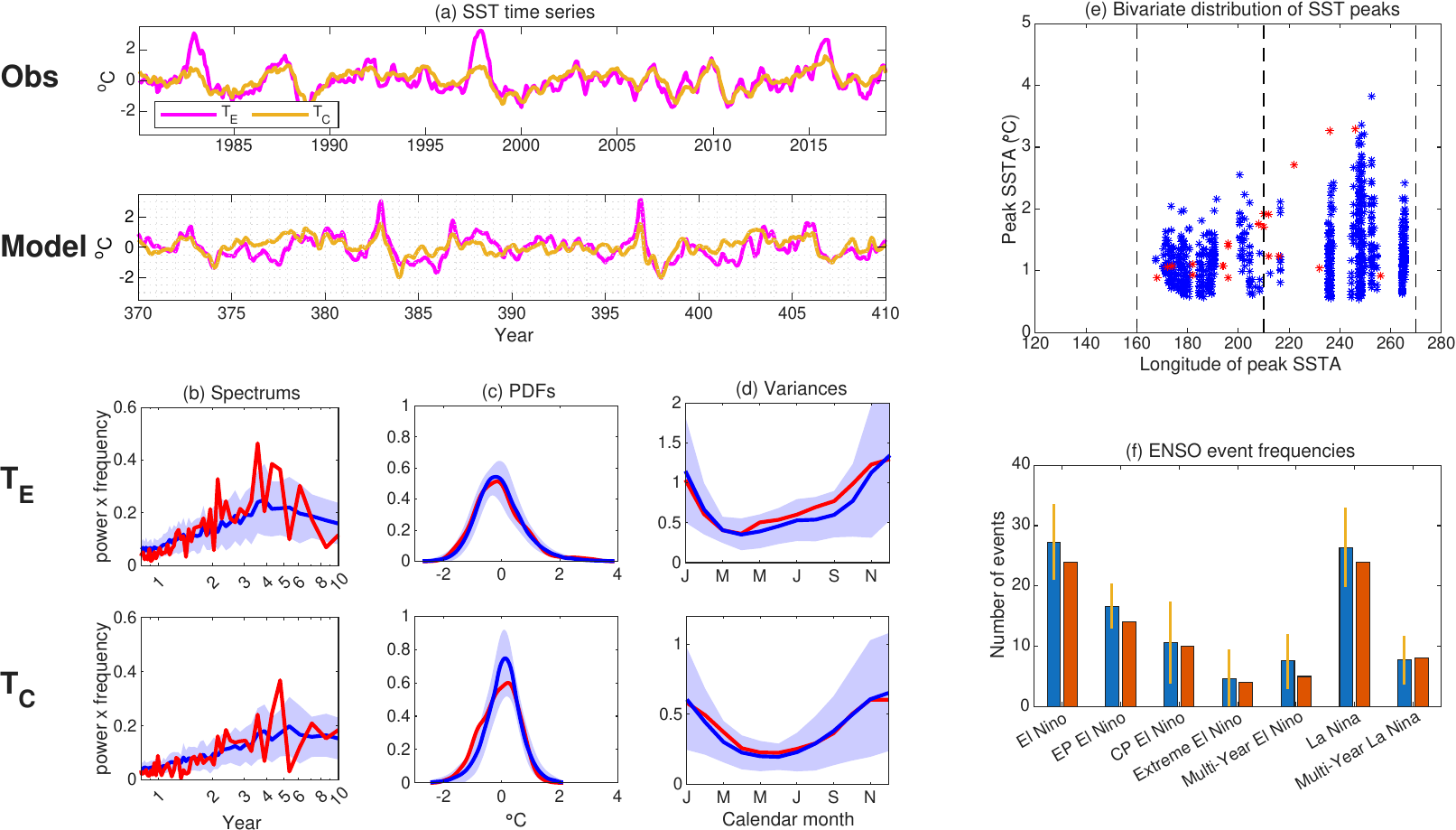}
\caption{Comparison of model simulations and statistical data with observational records. Observations cover the period from 1950 to 2019 (a total of 70 years). The model simulation spans 2,000 years. The power spectra, PDFs, and seasonal variances (Panels (b)--(d)) are computed over 50 non-overlapping 40-year blocks, whereas the ENSO event frequencies (Panel (f)) are computed over 28 non-overlapping 70-year segments, each matching the length of the observational record. Confidence intervals are estimated from these blocks.
Panel (a): Time series of $T_E$ and $T_C$.
Panels (b)–(d): Power spectra, PDFs, and variances as a function of months for $T_E$ and $T_C$, with shaded areas representing 95\% confidence intervals.
Panel (e): Bivariate distribution of DJF SST peaks.
Panel (f): Frequency of ENSO events over 70 years, with bars showing 95\% confidence intervals.
Model simulations are shown in blue, and observational data are shown in red.}\label{fig:Statistics_ENSO}
\end{figure}

Figure \ref{fig:Statistics_ENSO} presents the model simulations and associated statistics compared with observations. Panel (a) demonstrates that the simulated SST time series qualitatively match the observations. Notably, several large positive values are observed in $T_E$ (e.g., at $t=383$ and $397$), corresponding to strong EP El Ni\~no events. Similarly, there are periods (e.g., at $t=379$ and $399$) where $T_C$ exceeds 0.5$^\circ$C and is larger than $T_E$, indicating CP events. Panels (b)--(d) compare key statistical measures. First, the power spectra of the model simulations exhibit the dominant ENSO frequency between 3 and 5 years, reflecting the average oscillation period and its irregularity. However, the model underrepresents a secondary power peak between 2 and 3 years in the Ni\~no 3 SST spectrum. Additionally, the model accurately reproduces the non-Gaussian probability density functions (PDFs), capturing the positive skewness and one-sided fat tail of the Ni\~no 3 SST distribution. This allows the model to simulate extreme EP El Ni\~no events and the El Ni\~no-La Ni\~na asymmetry. Likewise, the model captures the negative skewness of the Ni\~no 4 SST distribution, preventing extreme events from occurring in the CP region. Furthermore, the model successfully reproduces the seasonal variation of SST variance, indicating a higher likelihood of event occurrence during boreal winter. Panel (e) shows the bivariate distribution of DJF SST peaks. Similar to observations, the strength of the SST peaks increases as events shift eastward in the Pacific, where CP events generally have weaker amplitudes than EP events in the model. Finally, Panel (f) displays the frequency of different ENSO events, which is a useful indicator of ENSO complexity. Overall, the model succeeds in reproducing the number of different ENSO events, consistent with nature. \ref{sec:additional_analysis} contains the statistics of other variables, the role of the decadal variability $I$, and the reconstructed spatiotemporal fields. Particularly, the intraseasonal wind statistics also match the observations, capturing the zero mean over time and, in a window centered on EP El Ni\~no events, a small positive shift in the mean leading up to and throughout these events (see Figure~\ref{fig:SI_ENSO_Time_Series}).

\begin{figure}[h]
\hspace*{-1.cm}\includegraphics[width=1.1\textwidth]{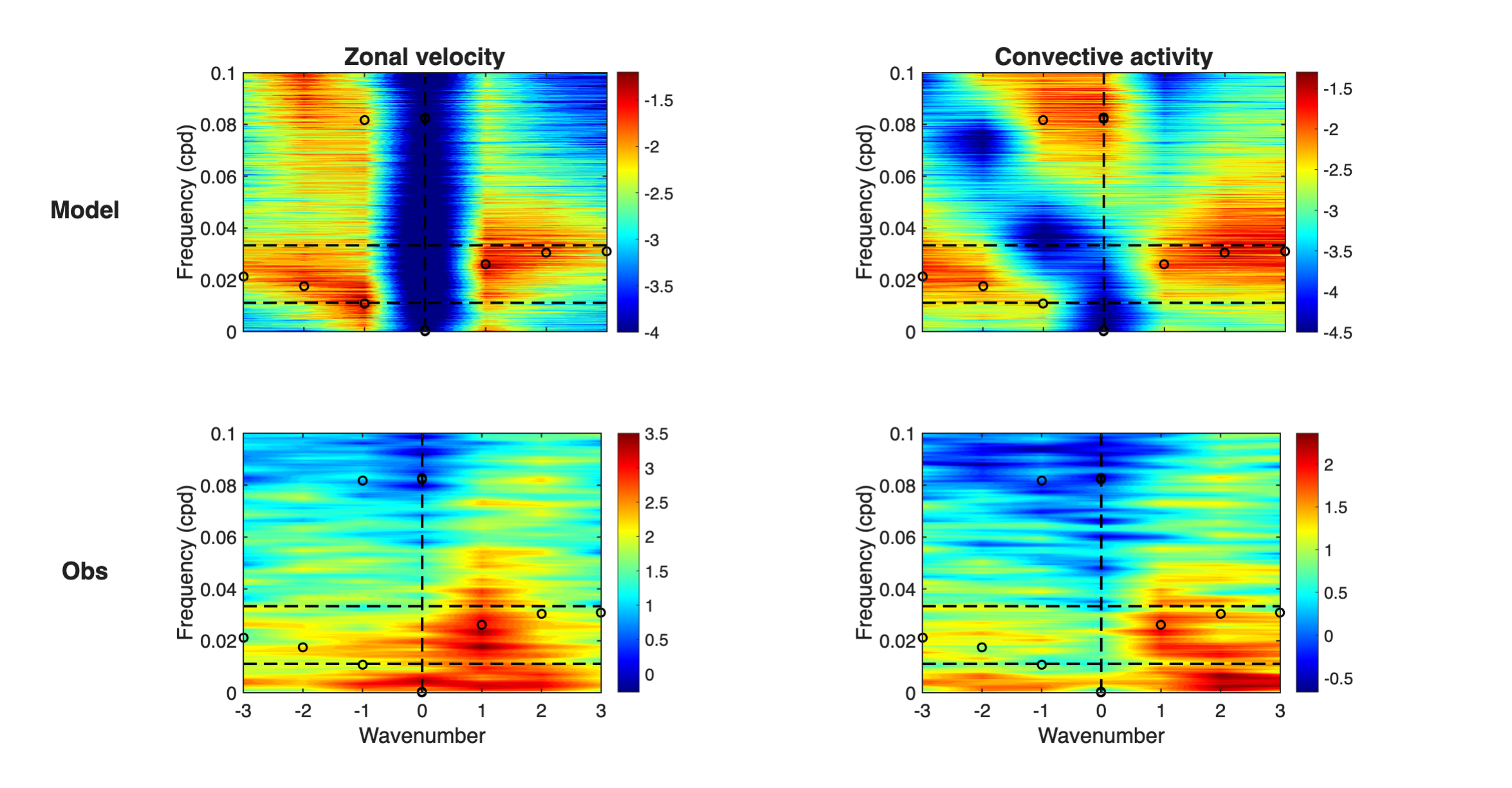}
\caption{Power spectra of atmospheric zonal velocity ($u$) and convective activity ($A$). The x-axis represents wavenumbers, while the y-axis shows frequency in cycles per day (cpd). As the conceptual model includes only the first three Fourier modes, the x-axis is limited to modes $k = \pm 3$. The top panels display results from the coupled model, and the bottom panels show observations. Black dots indicate the dispersion curves from linear analysis of the MJO skeleton model. The two dashed horizontal lines represent the intraseasonal band, spanning from 30 days (0.0333 cpd) to 90 days (0.0111 cpd).}\label{fig:Spectrum_MJO}
\end{figure}

The intraseasonal atmosphere reproduces the observed MJO statistics as well. Figure \ref{fig:Spectrum_MJO} presents the power spectra of atmospheric zonal velocity ($u$) and convective activity ($A$), two key variables for reconstructing the MJO. The x-axis spans wavenumbers from $k=-3$ to $k=3$, which are the wavenumbers used in the model. The y-axis represents frequency in cycles per day (cpd). The focus is on the intraseasonal band, highlighted between the two horizontal dashed lines. Black dots mark the dispersion curves from linear analysis of the MJO skeleton model. The high density within this band for modes $k=1, 2$, and $3$ indicates the dominant eastward-propagating MJO signal. Conversely, the westward-propagating moisture Rossby waves are captured by modes $k=-1,-2$, and $-3$. The power spectra of the model within the intraseasonal band resemble observations. On the other hand, the model underestimates the power density at lower frequencies (where cpd approaches zero). This is because the observations contain information across all temporal scales, whereas this model is primarily designed to capture intraseasonal variabilities. Further, the model density patterns at higher frequencies for modes $k=1, 2$, and $3$ are fairly consistent with observations. The observed MJO is predominantly eastward, whereas in the model the negative-wavenumber (westward) modes carry somewhat stronger spectral density than in observations, possibly due to the stochastic noise. Since the model only focuses on coupling MJO and ENSO, matching the higher frequency signals does not impact the model's function. Note that SST influences the intraseasonal atmosphere both through the background convective activity $\bar{A}$ and through the state-dependent moisture noise, the latter being the dominant statistical feedback that shapes the MJO--ENSO variability without directly shifting the mean intraseasonal state.

\begin{figure}[h]
    \centering
    \includegraphics[width=1\linewidth]{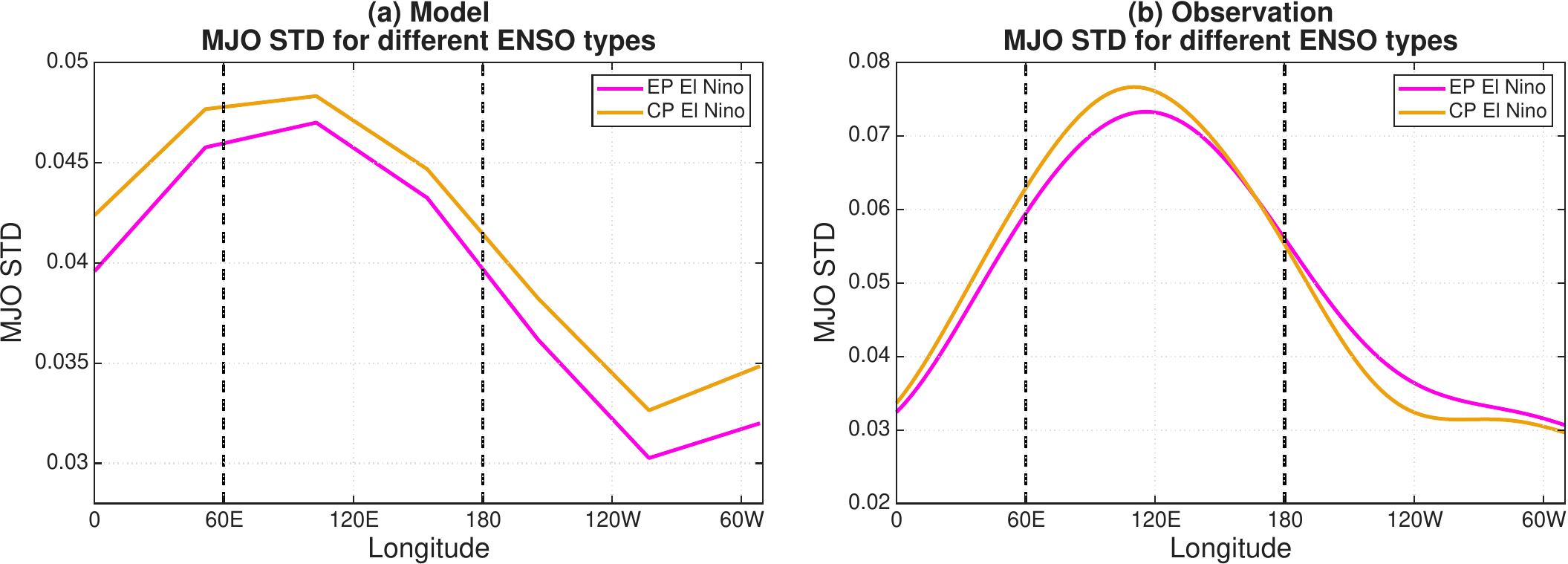}
    \caption{Standard deviation of MJO plotted as a function of longitude during EP El Ni\~no events (purple) and CP El Ni\~no events (orange). (a) displays the zonal standard deviation produced by the model and (b) shows the zonal standard deviation from observations. The dashed black lines mark the western boundary of the Indian Ocean and the eastern boundary of the WP.}
    \label{MJO_zonal_std}
\end{figure}

Figure \ref{MJO_zonal_std} presents the standard deviation of MJO as a function of longitude conditioned on CP and EP El Ni\~no events. The longitudes span from the prime meridian to $70^\circ W$. The warm pool region, which includes the Indian Ocean and the WP, is highlighted by dashed lines. The high MJO activity in this region indicates that the warm SSTs fuel the MJO. On the other hand, lower MJO variability outside this region reflects the suppression of MJO by cooler SSTs and the associated reduction in deep atmospheric convection. Note that the amplitude of the MJO signal in the model is slightly weaker than that of observations since the model is coarse grained and thus averages out the higher-frequency, larger-amplitude signals.

\subsection{Air-sea interactions}

\begin{figure}[h]
\hspace*{-0.3cm}\includegraphics[width=1\textwidth]{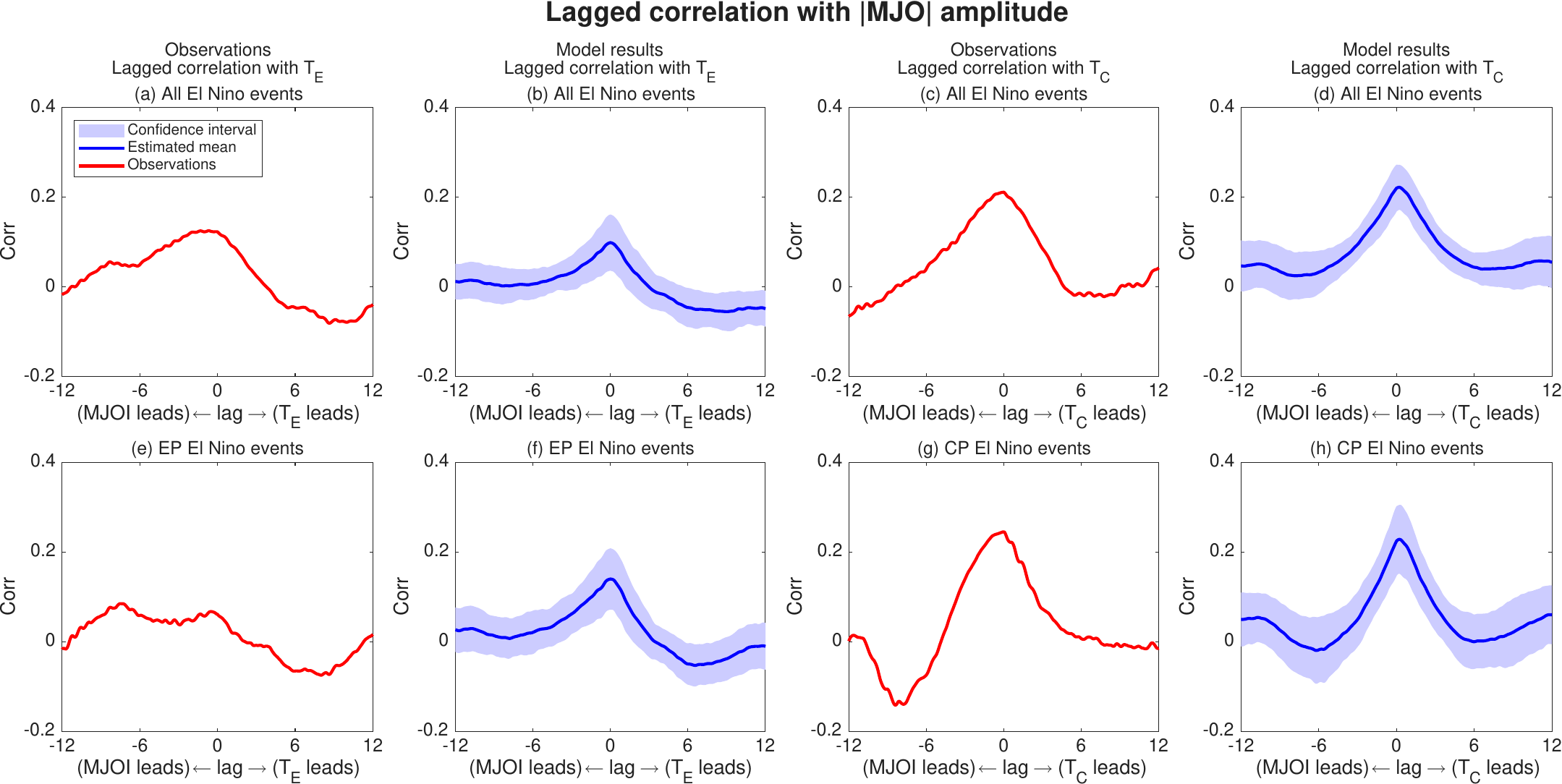}
\caption{Lagged correlation between the MJO and ENSO. Panels (a)--(b): Lagged correlation with $T_E$ for years when El Ni\~no events occur. Panels (c)--(d): lagged correlation with $T_C$ for the same years. Panels (e)--(f): Lagged correlation with $T_E$ for years with only EP El Ni\~no events. Panels (g)--(h): Lagged correlation with $T_C$ for years with only CP El Ni\~no events. On the x-axis, zero corresponds to the peak of the El Ni\~no event, with the unit of the x-axis being months. In each panel, the red curve denotes the observations, the blue curve the estimated mean of the model results, and the shaded area the confidence interval (one standard deviation across the model segments).}\label{fig:Lagged_Correlation}
\end{figure}

The air--sea coupling between the MJO and ENSO can be quantified through their lagged relationship. Figure \ref{fig:Lagged_Correlation} shows the lagged correlation between the MJO index strength $|$MJOI$|$ and an SST index, either $T_C$ or $T_E$. Since the MJO consists of seven time series corresponding to different points along the equator, we must introduce the index MJOI, which is the MJO signal averaged over the WP ($140^\circ$E–$160^\circ$W), to compute the lagged correlation. This correlation is calculated specifically for years when El Ni\~no events occur, aiming to highlight the statistical dependence between ENSO and the MJO. The analysis explores the lagged correlation across all El Ni\~no events, including both EP and CP types (Panels (a)--(d)), as well as conditioning separately on EP events (Panels (e)--(f)) and CP events (Panels (g)--(h)).

Overall, the lagged correlations from the model simulations align well with observations. The amplitude of the MJOI ($|$MJOI$|$) shows a clear correlation with SST, with the MJO becoming more intense prior to or during El Ni\~no phases. Notably, the lagged correlation with $T_E$ during all El Ni\~no phases exhibits asymmetry, with a longer correlation observed before an El Ni\~no event (Panel (a)). Such an extended correlation range reflects the process of MJO buildup and the triggering of El Ni\~no events. After the El Ni\~no peak, the MJO remains active for a few months until the SST anomaly dissipates. The model captures a similar, although less pronounced, lagged correlation pattern (Panel (b)). For $T_C$, the lagged correlation during all El Ni\~no events resembles that with $T_E$ in observations, though the model produces a slightly more symmetric correlation band. When conditioned only on CP events, the lagged correlation with $T_C$ is more symmetric in both observations and model results (Panels (g)--(h)), with significant correlations spanning a shorter window, from -5 to 5 months, compared to conditioning on all El Ni\~no events. Finally, Panels (e)--(f) depict the lagged correlation with $T_E$ conditioned on EP events. Here the model and observations have similar patterns, but the model correlations with $T_E$ are slightly stronger than observations. However, caution is needed when interpreting such a result since the number of EP events in the observational period is quite limited. Physically, this lagged asymmetry mirrors the two coupling pathways identified above. The correlation that leads the El~Ni\~no peak corresponds to the MJO wind stress forcing and triggering the ocean warming, whereas the correlation that lags the peak corresponds to the warm SST sustaining the MJO through the latent-heat feedback until the anomaly dissipates.

\subsection{Sensitivity tests}

\subsubsection{Impact of decadal variability}\label{subsec:decadal_mod}
\begin{figure}[h!]
    \centering
    \includegraphics[width=1\linewidth]{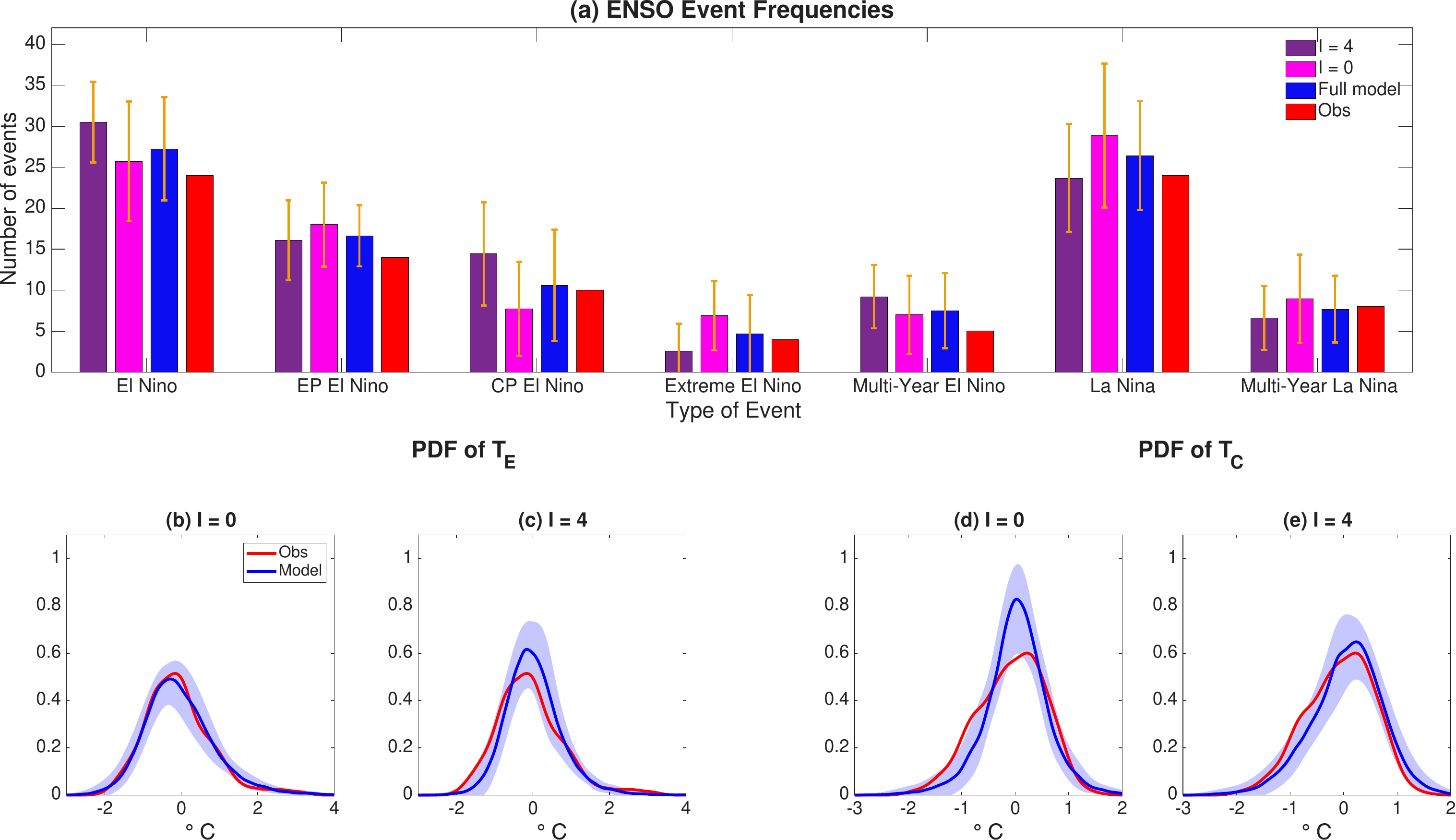}
    \caption{Comparison of the ENSO statistics from three model configurations. Panel (a) shows the ENSO event frequencies per 70 year period under different model regimes compared with observations. The results of the full model simulation are in blue, the model simulation with $I$ fixed at 0 are in pink, the model simulation with $I$ fixed at 4 are in purple, and observations are in red. Panels (b) and (c) show the PDF of $T_E$ while panels (d) and (e) show the PDF of $T_C$. Panels (b) and (d) show PDFs produced by the model where $I$ is fixed at 0, and panels (c) and (e) show that produced by model where $I$ is fixed at 4.}
    \label{fig:SI_ENSO_stats_I_exp}
\end{figure}

The decadal variable $I$ sets the background strength of the Walker circulation and thereby selects the predominant flavor of ENSO over a given epoch. To isolate its effect, Figure~\ref{fig:SI_ENSO_stats_I_exp} compares the ENSO statistics produced under three formulations of $I$: the full model, in which $I$ is driven by a simple stochastic process; the model fixed in the EP-dominant regime ($I=0$ in \eqref{Interannual_ocean_model}); and the model fixed in the CP-dominant regime ($I=4$). The decadal variable enforces these two regimes by modulating both the strength of the MJO wind forcing (the $\beta$s in \eqref{Interannual_ocean_model}) and the zonal advection in the CP (the $\rho I U$ term in \eqref{Interannual_ocean_model_T_C}).

Under the EP-dominant regime, the trade winds are weak, enhancing the influence of the highly variable MJO-related wind on the ocean dynamics, particularly in the EP. Consequently, $T_E$ has a larger variance (Panel (b)), and EP events are more frequent and intense (Panel (a); pink), producing more extreme non-Gaussian $T_E$ statistics. At the same time, the weak SST gradient suppresses zonal advection in the CP, decreasing the frequency of CP events (Panel (a); pink) and in turn the variance of $T_C$ (Panel (d)). The PDFs in Panels (b) and (d) reveal that the cold phase responds similarly, with more negative SST anomalies in the EP and weaker cooling in the CP; as a result, La Ni\~na events are somewhat more frequent in the EP-dominant regime. In contrast, under the CP-dominant regime the coupling coefficients are small, weakening the impact of the MJO-related wind and strengthening zonal advection in the CP. This decreases the variance of $T_E$ (Panel (c)) through less frequent and less intense EP events, while CP events become more frequent (Panel (a); purple), increasing the variance of $T_C$ (Panel (e)); the EP SST anomalies become more Gaussian (Panel (c)) and, despite the enhancement of CP cooling, the overall La Ni\~na frequency is slightly reduced. Thus $I$ shifts the entire oscillator spatial pattern rather than the warm phase alone. The corresponding time series of all model variables, together with an illustration of how $I$ modulates the SST Hovmoller diagram, are provided in Figure~\ref{fig:SI_ENSO_Time_Series}.

\subsubsection{Role of the coupling and the state-dependent noise}\label{subsec:coupling}
The coupled model links the MJO and ENSO through three channels, namely the MJO wind stress $\tau$ that forces the interannual ocean, the latent heat $E_q\propto T_C$ that feeds back on the atmosphere, and the two state-dependent noises, $\sigma_{\hat{A}_k}(Q,A)\dot{W}_{\hat{A}_k}$ in \eqref{Intraseasonal_atmosphere_model_A} and $\sigma_{\hat{Q}_k}(E_q)\dot{W}_{\hat{Q}_k}$ in \eqref{Intraseasonal_atmosphere_model_Q}. We probe each channel by switching it off in turn. In all comparisons that follow, the simulations use identical realizations of the stochastic forcing (the same random seed), so any difference reflects only the term being modified.

\begin{figure}[h!]
    \centering
    \includegraphics[width=1\linewidth]{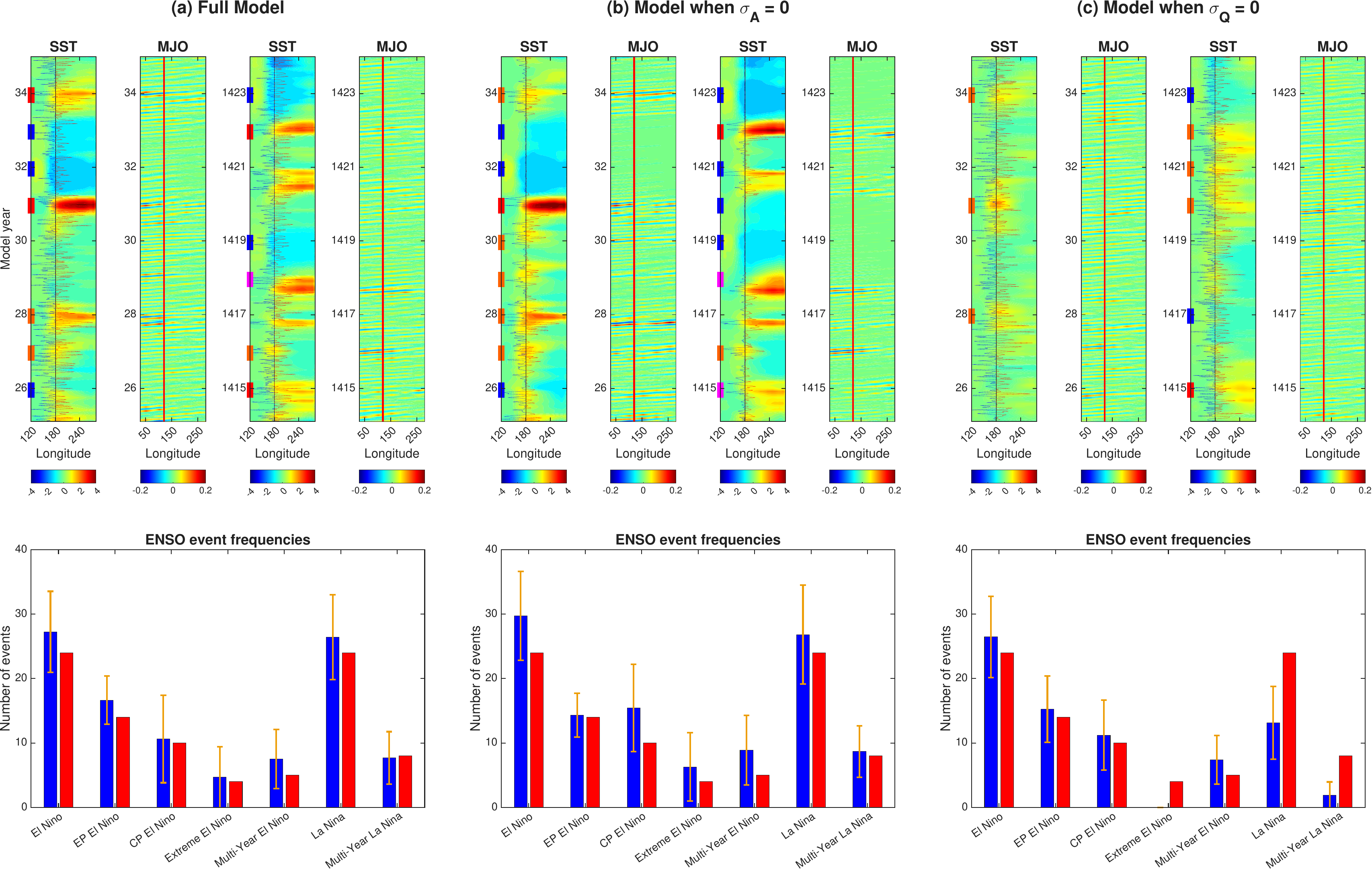}
    \caption{Sensitivity of the model to the state-dependent noises in the MJO skeleton. Panel (a): the full model. Panel (b): the model with $\sigma_{\hat{A}_k} = 0$, removing the state-dependent noise in the convective activity equation. Panel (c): the model with $\sigma_{\hat{Q}_k} = 0$, removing the state-dependent noise in the moisture equation. Each panel shows the corresponding SST ($^\circ$C) and MJO (unit-less) Hovmoller diagrams together with the ENSO event frequencies per 70 years compared with observations.}
    \label{fig:SI_Sensitivity_NoQnoise}
\end{figure}

We first examine the two state-dependent noises, which are essential for reproducing the MJO and its statistical coupling with ENSO (Figure~\ref{fig:SI_Sensitivity_NoQnoise}). The convective-activity noise $\sigma_{\hat{A}_k}$ generates the intermittency of the MJO. Without it ($\sigma_{\hat{A}_k}=0$; Panel (b)), MJO events become more regular and the MJO--ENSO relationship becomes overly deterministic, producing a much larger lagged correlation than observed, while the ENSO event statistics remain close to the full model. The moisture noise $\sigma_{\hat{Q}_k}$ carries the statistical feedback from the latent heat, allowing warm SST to strengthen the MJO and to favor westerly wind bursts ahead of EP El Ni\~no events \cite{puy2016modulation}. Without it ($\sigma_{\hat{Q}_k}=0$; Panel (c)), the MJO weakens and strong El Ni\~no events become less frequent, as seen in both the Hovmoller diagrams and the ENSO event frequencies. In both cases the SST--MJO lagged correlation is biased away from observations.

\begin{figure}[h]
\hspace*{-0.75 cm}\includegraphics[width=1.0\textwidth]{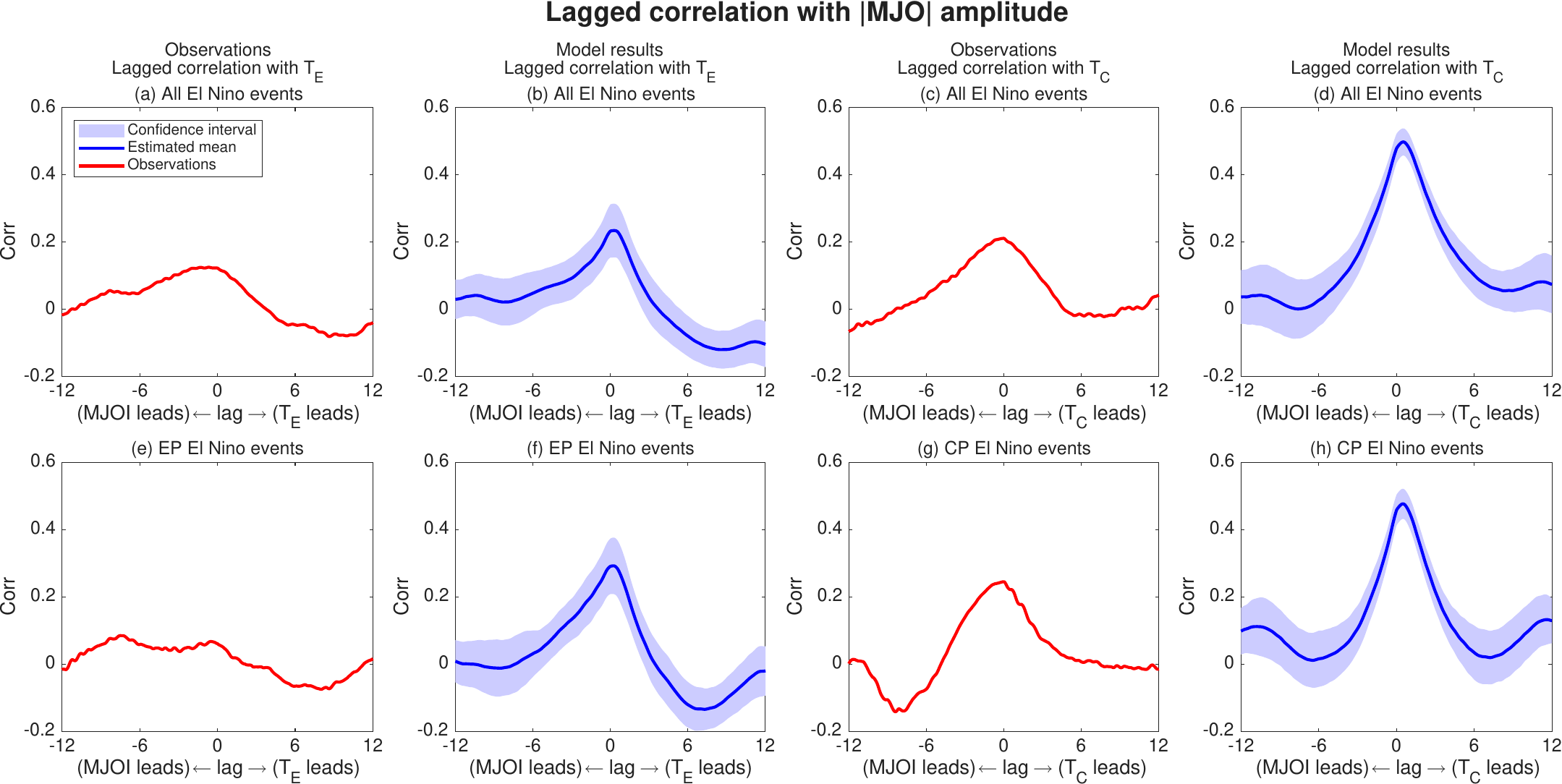}
\caption{Lagged correlation between the MJO and ENSO, similar to Figure \ref{fig:Lagged_Correlation}, but for the system with no state-dependent noise in the convective activity equation ($\sigma_{\hat{A}_k} = 0$).}
\label{fig:Lagged_Correlation_all_withoutA}
\end{figure}

Figure \ref{fig:Lagged_Correlation_all_withoutA} presents the lagged correlation between MJO and ENSO when the state-dependent noise in convective activity is removed ($\sigma_{\hat{A}_k} = 0$), based on the MJOI and SST time series. Compared with the full model results, as shown in Figure \ref{fig:Lagged_Correlation}, the lagged correlation without this noise is higher than observed. As a result, the relationship between MJO and ENSO becomes more deterministic, deviating from the natural system. The findings here highlight the crucial role of state-dependent noise in convective activity, which is essential for generating the many intermittent MJO events seen in nature.

\begin{figure}[h]
\hspace*{-0.75 cm}\includegraphics[width=1.0\textwidth]{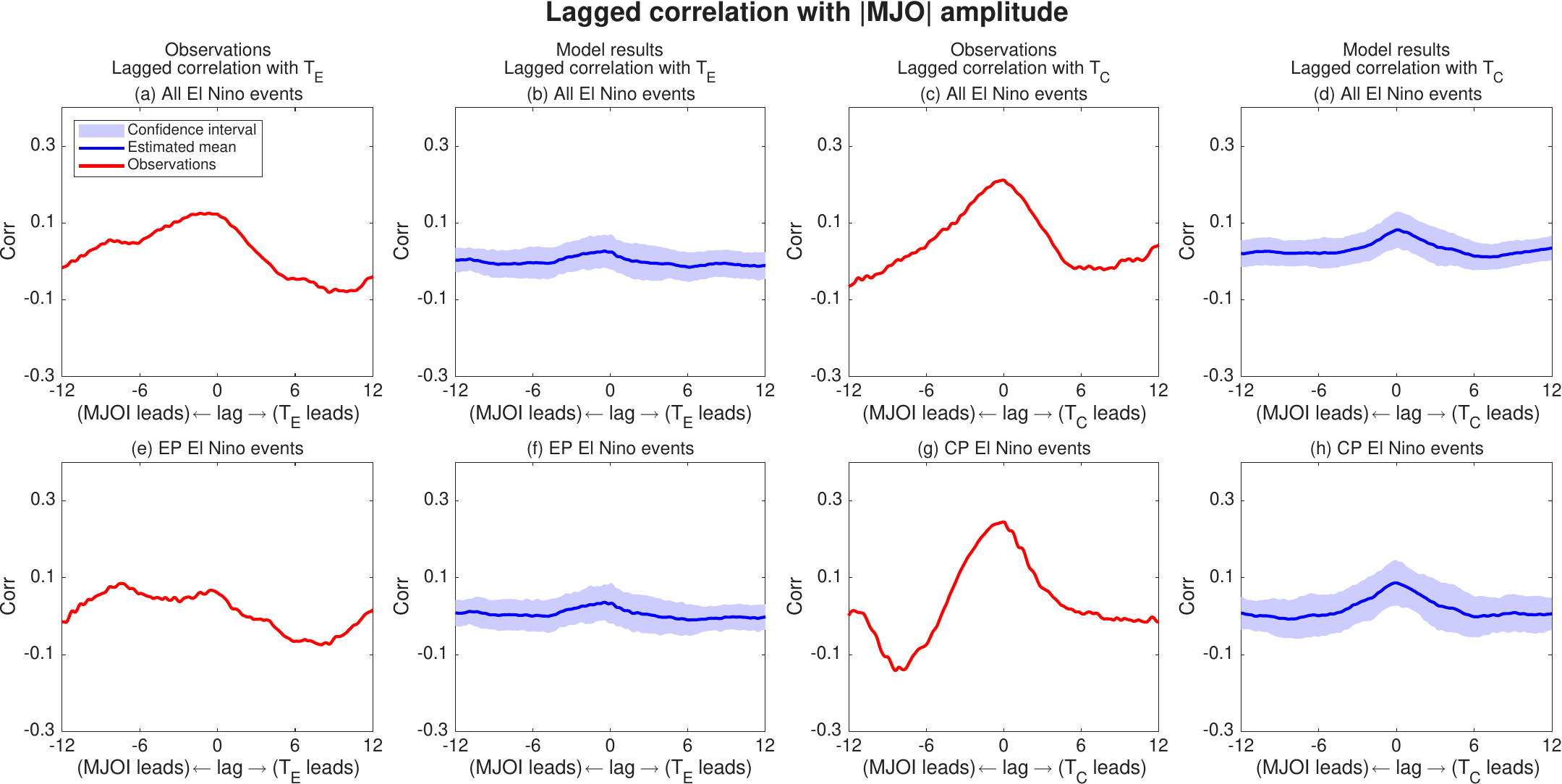}
\caption{Lagged correlation between the MJO and ENSO, similar to Figure \ref{fig:Lagged_Correlation}, but for the system with no state-dependent noise in the low-level moisture equation ($\sigma_{\hat{Q}_k} = 0$).}
\label{fig:Lagged_Correlation_all_withoutQ}
\end{figure}

Figure \ref{fig:Lagged_Correlation_all_withoutQ} depicts the lagged correlation between MJOI and SST time series for the model with no noise in the low-level moisture equation. In comparison with observations, and the full model results (Figure \ref{fig:Lagged_Correlation}), without this noise the link between MJO and SST is weakened. Removing this state dependent noise breaks the feedback loop between the atmospheric and oceanic components. Ultimately wind bursts are reduced thus inhibiting the formation of extreme El Ni\~no events and in turn reducing MJO activity.

\begin{figure}[h!]
    \centering
    \includegraphics[width=\linewidth]{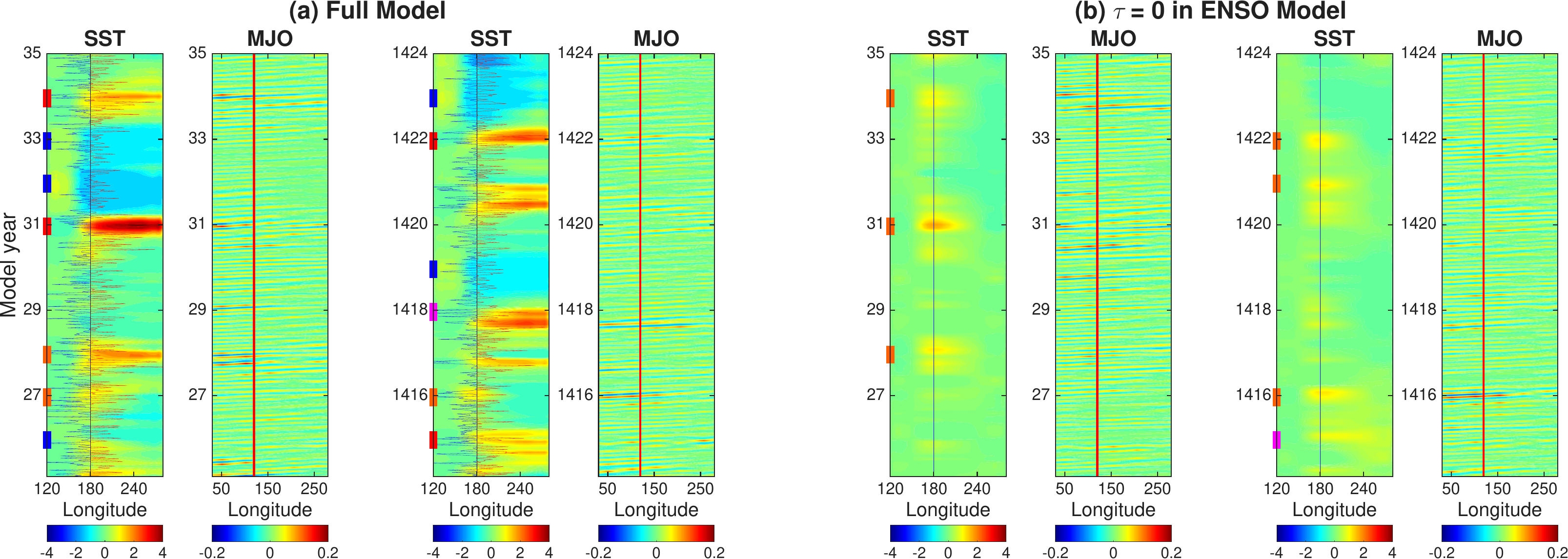}
    \caption{SST ($^\circ$C) and MJO (unit-less) Hovmoller diagrams produced by two model configurations: the full model (Panel (a)) and the model with the MJO wind forcing removed from the interannual ocean equations ($\tau = 0$; Panel (b)).}
    \label{hov_no_tau}
\end{figure}

Next, removing the MJO wind forcing entirely ($\tau = 0$; Figure~\ref{hov_no_tau}) strips the interannual ocean of the high-frequency stochastic triggering supplied by the MJO. Because the intraseasonal wind, through the multiplicative noise in the moisture process, tends to favor westerly wind bursts ahead of EP El Ni\~no events, removing it weakens EP warming and lowers the frequency of extreme warm events. This confirms the role of the MJO wind in sustaining the ocean--atmosphere feedback loop that underlies ENSO.

\begin{figure}[h!]
    \hspace*{-1cm}\includegraphics[width=1.05\textwidth]{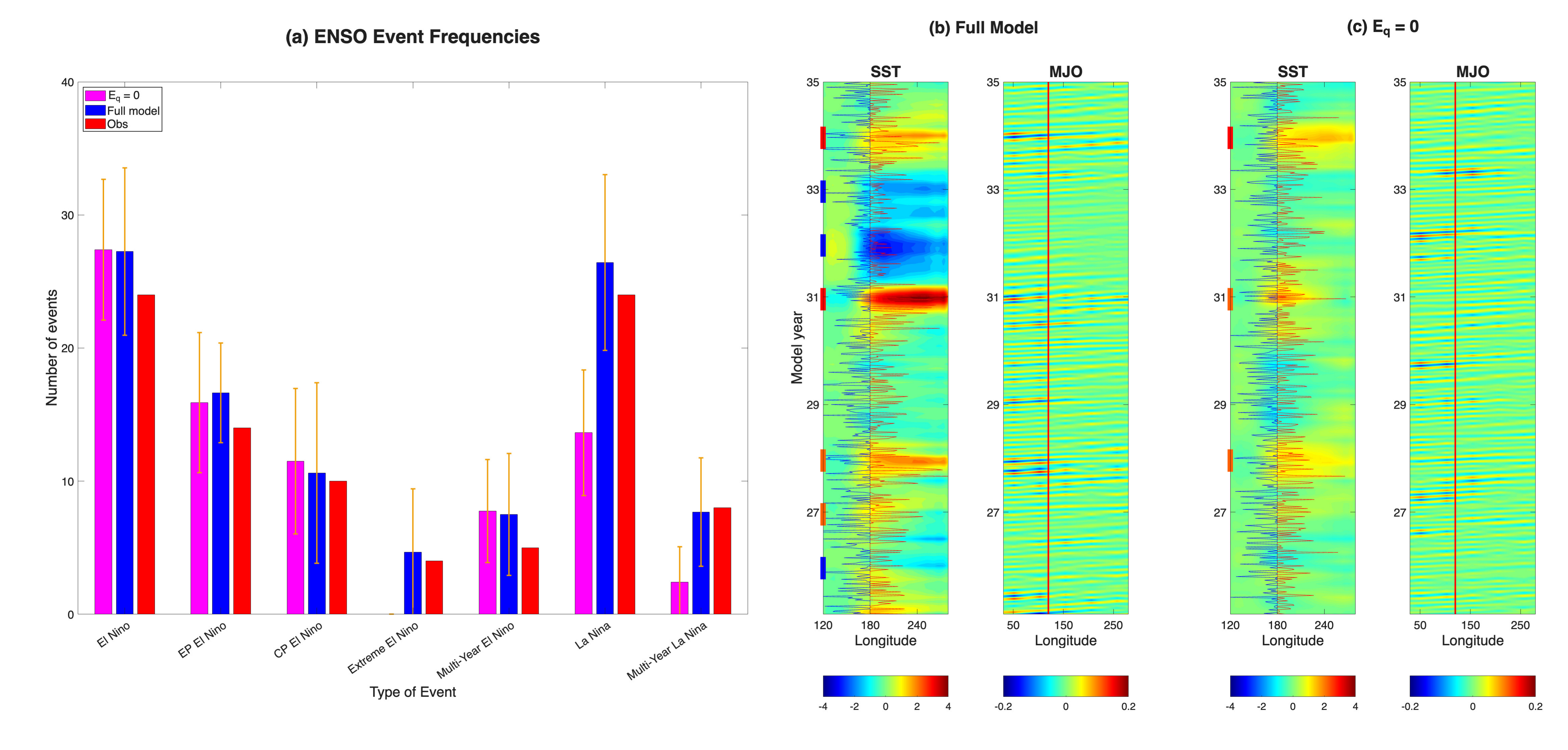}
    \caption{Comparison of the full model and the model with the latent heat set to zero ($E_q = 0$). Panel (a) shows the ENSO event frequencies per 70 year period; the full model results are in blue, the results for $E_q = 0$ in pink, and observations in red. Panels (b) and (c) show the Hovmoller diagrams of SST ($^\circ$C) and MJO (unit-less) over the same period, produced by the full model and by the $E_q = 0$ model, respectively.}
    \label{fig:SI_event_freq_Eq_constant}
\end{figure}

Finally, setting the latent heat to zero ($E_q=0$; Figure~\ref{fig:SI_event_freq_Eq_constant}) removes the SST control of the MJO altogether, since ENSO influences the MJO through the latent heat that enters the multiplicative noise of the moisture process ($\sigma_Q$ in \eqref{Intraseasonal_atmosphere_model_Q}). When $E_q=0$, the wind no longer responds to SST anomalies, so the warming potential in the EP is limited: EP events can still be triggered, but only when the atmospheric and oceanic conditions align by chance, and they therefore decrease in frequency and intensity (Panel (a); pink). The weakened EP warming steepens the thermocline and strengthens the zonal current, boosting zonal advection and increasing the frequency of CP events; without the SST feedback the MJO signal is also weakened and loses its clear relation to ENSO (Panel (c)). Among all channels, removing the latent heat has the largest effect on the coupling. Taken together, these experiments quantify how each dynamical and stochastic channel contributes to ENSO diversity, extreme-event frequency, and the MJO--ENSO correlation, and demonstrate that the state-dependent noise is as essential as the deterministic wind forcing for reproducing the observed coupled behavior.

\section{Conclusion and discussions}\label{Sec:Conclusion}\label{Sec:Discussion}

In this paper, we develop a stochastic conceptual model to describe the large-scale dynamics of the coupled ENSO and MJO systems. The interannual ocean is represented by a three-box recharge oscillator spanning the equatorial Pacific, the intraseasonal atmosphere by a low-order stochastic MJO skeleton model, and the slow background state by a simple decadal process. State-dependent noise is introduced alongside the wind forcing and latent heat to enhance the multi-scale statistical interactions and to improve the simulation of extreme events. The model reproduces the observed diversity and non-Gaussian statistics of ENSO, including the distinct EP and CP events at realistic frequencies, together with the energy spectra of both ENSO and the MJO and the seasonal phase locking of ENSO. It also captures the two-way air-sea interaction, in which intraseasonal wind bursts precede warm events and the MJO convection extends eastward toward the eastern Pacific during El Ni\~no. A systematic set of mechanism experiments then clarifies the distinct contribution of each dynamical and stochastic channel to this coupled behavior.

The present model builds on two established lines of conceptual modeling. Like the multiscale stochastic model of \citeA{chen2022multiscale}, it couples a recharge-oscillator ocean to a stochastic atmosphere. In the present model, the wind is generated by a low-order MJO skeleton, so that the wind forcing carries intraseasonal, eastward-propagating structure and can respond to the ocean state. Like the stochastic skeleton model of \citeA{thual2014stochastic}, it represents the MJO through the skeleton dynamics. The latent-heat feedback and the state-dependent moisture noise introduced here further let the ocean modulate the MJO, closing the two-way ENSO--MJO loop. In this way, the model offers an explicit, dynamically consistent bridge between the intraseasonal and interannual scales \cite{vialard2024nino}.

Because the model is computationally inexpensive and statistically accurate, it may be useful in the future as a component in multi-model data assimilation, supplying dynamically consistent ENSO and MJO statistics at low cost to improve reanalysis and the initialization of more complex models \cite{parsons2021multi, bach2023multi, chen2019multi}. Because the coupled model resolves the MJO explicitly rather than as prescribed noise, it also provides a controlled setting for predictability studies. In particular, it can be used to quantify how the MJO affects ENSO forecast skill. It can also test whether explicitly resolving the MJO is necessary, or whether a simplified wind parameterization suffices for prediction. Its low computational cost further makes it well suited to generating the very long simulations needed to study rare and extreme events, to characterize the statistical response of the coupled system to climate perturbations, and to produce training data for machine-learning approaches \cite{zhang2024physics}. The decadal variable further gives the model the potential to project the large-scale patterns of the MJO and ENSO under global-warming scenarios, using predicted changes in Walker-circulation strength. More broadly, the explicit and adaptable coupling structure makes the model a convenient platform for testing competing theories of ENSO and MJO dynamics and the roles of nonlinearity and noise therein.

\section*{Open Research}
The code used to process the data and create the figures was written in MATLAB. The code and output data of the experiments are available on Zenodo \cite{zenodo1,zenodo2}. The observational datasets are publicly available: GODAS and ERSSTv5 sea surface temperature, NOAA interpolated OLR, and NCEP--NCAR reanalysis (see Section~\ref{Sec:Datasets} for sources).

\acknowledgments
The research of N.C. is funded by the Office of Naval Research N00014-24-1-2244. Y.Z. is partially supported as a research assistant under this grant. C.M. is partially supported as a research assistant by the aforementioned ONR award and partially by the University of Wisconsin--Madison Office of the Vice Chancellor for Research, with funding from the Wisconsin Alumni Research Foundation. The authors declare no competing interests. N.C. conceived and supervised the study. Y.Z. and C.M. developed the model, performed the simulations, and analysed the results. All authors contributed to interpreting the results and writing and reviewing the manuscript. The authors thank the editors and reviewers for their constructive comments.

\appendix
\section{A brief overview of the stochastic skeleton model for the MJO}\label{sec:skeleton_overview}
The MJO skeleton model is a nonlinear oscillator model for the MJO skeleton as a neutrally stable wave \cite{majda2009skeleton, majda2011nonlinear}. The core mechanism behind this oscillation is the interaction between (i) planetary-scale lower-tropospheric moisture anomalies, denoted by $q$, and (ii) subplanetary-scale convection and wave activity, represented by $a$. The planetary envelope, $a \geq 0$, characterizes the collective influence of unresolved subplanetary convection and wave activity. A critical aspect of the $q$-$a$ interaction lies in the assumption that moisture anomalies $q$ directly affect the rate of change of $a$, expressed by the relation $a_t = \Gamma q a$, where $\Gamma > 0$ is a constant governing the strength of this interaction.

In the skeleton model, the interaction between moisture anomalies ($q$) and convection/wave activity ($a$) is further coupled with the linearized primitive equations projected onto the first vertical baroclinic mode. The non-dimensionalized system of equations is given as follows:
\begin{subequations}\label{Skeleton_Model}
\begin{align}
    u_t - yv - \theta_x &= 0,\label{Skeleton_Model_U}\\
    yu - \theta_y &= 0,\label{Skeleton_Model_V}\\
    \theta_t - u_x - v_y &= \bar{H}a - s^{\theta},\label{Skeleton_Model_Theta}\\
    q_t + \bar{Q} (u_x + v_y) &= -\bar{H}a + s^q,\label{Skeleton_Model_q}\\
    a_t &= \Gamma q a,\label{Skeleton_Model_a}
\end{align}
\end{subequations}
with periodic boundary conditions along the equatorial belt and planetary-scale equatorial long-wave scaling. In the dry dynamics (\eqref{Skeleton_Model_U}–\eqref{Skeleton_Model_Theta}), $u$, $v$, and $\theta$ represent zonal velocity, meridional velocity, and potential temperature, respectively, while \eqref{Skeleton_Model_q} governs the evolution of low-level moisture ($q$). All variables are anomalies from a radiative-convective equilibrium, except for $a$. The skeleton model incorporates a minimal set of parameters: $\bar{Q}$ represents the background vertical moisture gradient, and $\Gamma$ is a proportionality constant for the $q$-$a$ interaction. While $\bar{H}$ is not dynamically influential, it defines a cooling/drying rate $\bar{H}a$ in dimensional terms. External cooling and moistening effects, $s^\theta$ and $s^q$, must be prescribed to complete the system.

Next, the system \eqref{Skeleton_Model} is projected onto the first Hermite function in the meridional direction, where the fields are represented as  $a(x,y,t) = \tilde{A}(x,t)\phi_0, q = Q\phi_0, s^q = S^q\phi_0, s^\theta = S^\theta\phi_0$, where $\phi_0(y) = \sqrt{2}(4\pi)^{-1/4}\exp(-y^2/2)$.
This choice of meridional heating structure excites only Kelvin waves and the first symmetric equatorial Rossby waves. The resulting meridionally truncated system is:
\begin{subequations}\label{Skeleton_Model_Proj}
\begin{align}
    K_t + K_x &= (S^\theta - \bar{H}\tilde{A})/2,\label{Skeleton_Model_Proj_K}\\
    R_t - R_x/3 &= (S^\theta - \bar{H}\tilde{A})/3,\label{Skeleton_Model_Proj_R}\\
    Q_t + \bar{Q}(K_x - R_x/3) &= (\bar{H}\tilde{A} - S^q)(\bar{Q}/6-1),\label{Skeleton_Model_Proj_Q}\\
    \tilde{A}_t &= \Gamma Q\tilde{A},\label{Skeleton_Model_Proj_A}
\end{align}
\end{subequations}
provided that $S^\theta=S^q$. Note that $\tilde{A}$ represents the total convective activity, namely the summation of the anomaly $A$ and the background state $\bar{A}$. The dry dynamics components can be reconstructed as follows:
\begin{equation}\label{DryComponents}
\begin{split}
    u &= (K - R)\phi_0 + R\phi_2/\sqrt{2},\\
    v &= (4\partial_xR - \bar{H}\tilde{A}) \phi_1 /3\sqrt{2},\\
    \theta &= -(K + R)\phi_0 - R\phi_2/\sqrt{2}.
\end{split}
\end{equation}
Here, the higher-order Hermite functions $\phi_1(y)=2y(4\pi)^{-1/4}\exp(-y^2/2)$ and $\phi_2(y)=(2y^2-1)(4\pi)^{-1/4}\exp(-y^2/2)$, though irrelevant to the dynamics, are necessary for retrieving the MJO's quadrupole structure \cite{majda2009skeleton}.

The original stochastic skeleton model \cite{thual2014stochastic} introduces a simple stochastic parameterization to represent synoptic-scale processes. In this model, the amplitude equation \eqref{Skeleton_Model_a} or \eqref{Skeleton_Model_Proj_A} is replaced by a stochastic birth-death process, which allows for intermittent fluctuations in the synoptic activity envelope \cite{gardiner2009stochastic}. Let $\tilde{A}$ be a random variable taking discrete values $\tilde{A} = \Delta \tilde{A} \eta$, where $\eta$ is a positive integer. The transition probabilities between states over a time step $\Delta t$ are given by:
\begin{equation}\label{ProbTrans}
\begin{split}
    &P\{\eta(t+\Delta t)=\eta(t)+1\} = \lambda\Delta t+o(\Delta t),\\
    &P\{\eta(t+\Delta t)=\eta(t)-1\} = \mu\Delta t+o(\Delta t),\\
    &P\{\eta(t+\Delta t)=\eta(t)\} = 1-(\lambda+\mu)\Delta t+o(\Delta t),\\
    &P\{\eta(t+\Delta t)\neq\eta(t)+1,\eta(t),\eta(t)-1\} = o(\Delta t),
\end{split}
\end{equation}
where $\lambda$ and $\mu$ are the birth and death rates defined as:
\begin{equation}\label{lambda_mu}
    \lambda =
    \begin{cases}
    \Gamma|q|\eta + \delta_{\eta0}\qquad& \mbox{if~} q\geq0\\
    \delta_{\eta0}\qquad& \mbox{if~} q<0
    \end{cases}\qquad\mbox{and}\qquad
    \mu =
    \begin{cases}
    0\qquad& \mbox{if~} q\geq0\\
    \Gamma|q|\eta\qquad& \mbox{if~} q<0
    \end{cases}
\end{equation}
with $\delta_{\eta0}$ being the Kronecker delta operator. These transition rates ensure that $\partial_tE(\tilde{A}) = \Gamma E(Q\tilde{A})$ for $\Delta \tilde{A}$ small, recovering the average $Q$-$\tilde{A}$ interaction described in \eqref{Skeleton_Model}.

In \cite{chen2016filtering}, a continuous stochastic differential equation (SDE) for convective activity is derived for small $\Delta a$:
\begin{equation}\label{Approx_a}
    \frac{\d \tilde{A}}{\d t} = \Gamma Q \tilde{A}  +  \sqrt{\Delta \tilde{A}\Gamma |Q| \tilde{A}}\dot{W}_A.
\end{equation}
where $\dot{W}_A$ represents a standard Wiener process, capturing the stochastic fluctuations in convective activity.

The leading three Fourier modes from the system \eqref{Skeleton_Model_Proj_K}–\eqref{Skeleton_Model_Proj_Q}, along with the stochastic equation \eqref{Approx_a}, are employed to construct a conceptual model for the atmospheric intraseasonal component. It aims to capture the essential large-scale dynamics and provide a simplified yet effective representation of intraseasonal atmospheric processes.

\section{Data processing and spatiotemporal reconstruction}

\subsection{Processing the observational data for the intraseasonal atmospheric model}\label{sec:data_processing}
The processing of observational data for the intraseasonal atmospheric model follows the procedure outlined in \cite{stechmann2015identifying}.

\noindent\textbf{Zonal velocity $u$}: The zonal velocity data set contains values at different layers in the vertical direction. The MJO skeleton model takes into account only the first baroclinic mode of $u$, which is defined by the following expression:
\begin{equation}
u:=u_{B C, 1}=\frac{u(850 \mathrm{hPa})-u(200 \mathrm{hPa})}{2 \sqrt{2}}.
\end{equation}

\noindent\textbf{Geopotential temperature $\theta$}: The geopotential temperature dataset, denoted as $\theta$, is related to the geopotential height $Z$, which is also measured at various vertical layers. The first baroclinic mode of $\theta$ is expressed as follows:

\begin{equation}
\theta:=\theta_{B C, 1}=-Z_{B C, 1}=-\frac{Z(850 \mathrm{hPa})-Z(200 \mathrm{hPa})}{2 \sqrt{2}}.
\end{equation}

The reference scales $\theta$ and $Z$ are approximately $\bar{\alpha} \approx 15.6 \mathrm{~K}$ and $c^2 / g \approx 265 \mathrm{~m}$, respectively. To achieve non-dimensionalization, the geopotential height data $Z$ is divided by $c^2 / g$.

\noindent\textbf{Moisture $Q$}: The variable $Q$ represents the lower tropospheric anomaly of water vapor near 850 $\mathrm{hPa}$ and is defined as follows:
\begin{equation}
Q=\frac{1}{4} q(925 h P a)+\frac{1}{2} q(850 h P a)+\frac{1}{4} q(725 h P a).
\end{equation}

Note that in the datasets for $q$, data at the 725 hPa level is unavailable; therefore, the data from the 700 hPa level is used instead. Additionally, $Q$ has been non-dimensionalized using the natural reference scale $L_v / c_p \bar{\alpha}$, where $\bar{\alpha}$ represents the reference potential temperature scale.

\noindent\textbf{Convective activity $A$}: The outgoing longwave radiation (OLR) is used as a surrogate for $A$ and is expressed by the following relationship:
\begin{equation}
\bar{H} A = -H_{\mathrm{OLR}} \times \mathrm{OLR},
\end{equation}
where $H_{\mathrm{OLR}}=0.06 \mathrm{Kday}^{-1}\left(\mathrm{Wm}^{-2}\right)^{-1}$ is an estimated constant.

Using the meridional basis functions, the definitions of the Kelvin wave $K$ and the first symmetric equatorial Rossby wave $R$ are given by:
\begin{equation}\label{Eq:KR}
\begin{aligned}
& K=\frac{1}{2}\left(u_0-\theta_0\right) \text { and } \\
& R=-\frac{1}{4}\left(u_0+\theta_0\right)+\frac{\sqrt{2}}{4}\left(u_2-\theta_2\right),
\end{aligned}
\end{equation}
where $u_m$ and $\theta_m$ represent the meridional projections.

\subsection{Reconstruction of the MJO from the characteristic variables}\label{sec:MJO_reconstruct}

The structure of the MJO can be characterized by the eigenvector associated with the linearized form of \eqref{Skeleton_Model_Proj}. Let $\mathbf{U} = (K, R, Q, A)^\mathtt{T}$ represent the collection of state variables in \eqref{Skeleton_Model_Proj}. By assuming a plane-wave ansatz for $\mathbf{U}$ and substituting it into \eqref{Skeleton_Model_Proj}, we derive an eigenvalue problem for each wavenumber $k$. This results in a four-dimensional system that features four eigenmodes: dry Kelvin, moisture Kelvin, dry Rossby, and MJO. Each of these waves can be expressed as a linear combination of $K, R, Q$ and $A$.

Let the eigenvalue corresponding to the MJO for wavenumber $k$ be denoted as follows:
\begin{equation*}
    \omega_k := \omega_{\mbox{\tiny MJO}}(k).
\end{equation*}

The evolution of the MJO for wavenumber $k$ is then represented by the time series of the Fourier coefficients $\hat{K}_k, \hat{R}_k, \hat{Q}_k$, and $\hat{A}_k$ projected onto the eigenvector associated with $\omega_k$. The numerical values of the eigenvalues and eigenvectors pertaining to the MJO modes are presented in Table \ref{Table_MJO_eigenmodes}.

\begin{table*}[h]
\begin{center}
\begin{tabular}{lccccc}
  \hline
  $k$ & $\omega_{\mbox{\tiny MJO}}(k)$ & $\hat{K}$ & $\hat{R}$ & $\hat{Q}$ & $\hat{A}$ \\\hline
  $1$ & $0.0261$ & $0.4577i$ & $-0.4088i$ & $-0.2513i$ & $0.7485$ \\
  $2$ & $0.0304$ & $0.2343i$ & $-0.2857i$ & $-0.3389i$ & $0.8652$ \\
  $3$ & $0.0309$ & $0.1537i$ & $-0.2177i$ & $-0.3561i$ & $0.8956$ \\
  \hline
\end{tabular}
\end{center}
\caption{Eigenvalues $\omega_{\mbox{\tiny MJO}}(k)$ and eigenvectors $\hat{\mathbf{e}}_{\mbox{\tiny MJO}}(k)$ for the MJO skeleton are shown for zonal wavenumbers $k=1, 2$, and $3$. The frequencies are presented in units of cycles per day (cpd). The model parameters are listed in Table \ref{params}.}\label{Table_MJO_eigenmodes}
\end{table*}

It is worth noting that the frequencies of the first three modes remain nearly constant. The final step in constructing the MJO signal involves summing the modes for
$k=1,2$, and $k=3$, followed by the application of a temporal filter to retain only the signals within the intraseasonal band, specifically between 30 and 90 days.

\subsection{Reconstruction of the ENSO spatiotemporal patterns using bivariate regression}\label{sec:spatial_reconstruct}

Recall that the stochastic conceptual model includes two SST variables, $T_C$ and $T_E$, as defined in \eqref{Interannual_ocean_model}. These variables can be used to approximately reconstruct the entire SST field across the equatorial Pacific. The method employed here is a bivariate regression,
\begin{equation}\label{Reconstruction_ENSO}
\mbox{SST}(x,t) = a_E(x) T_E(t) + a_C(x)T_C(t),
\end{equation}
where $a_E(x)$ and $a_C(x)$ are the spatial basis functions that depend on the location $x$.

\begin{figure}[h]
\centering
\hspace*{-0cm}\includegraphics[width=0.65\textwidth]{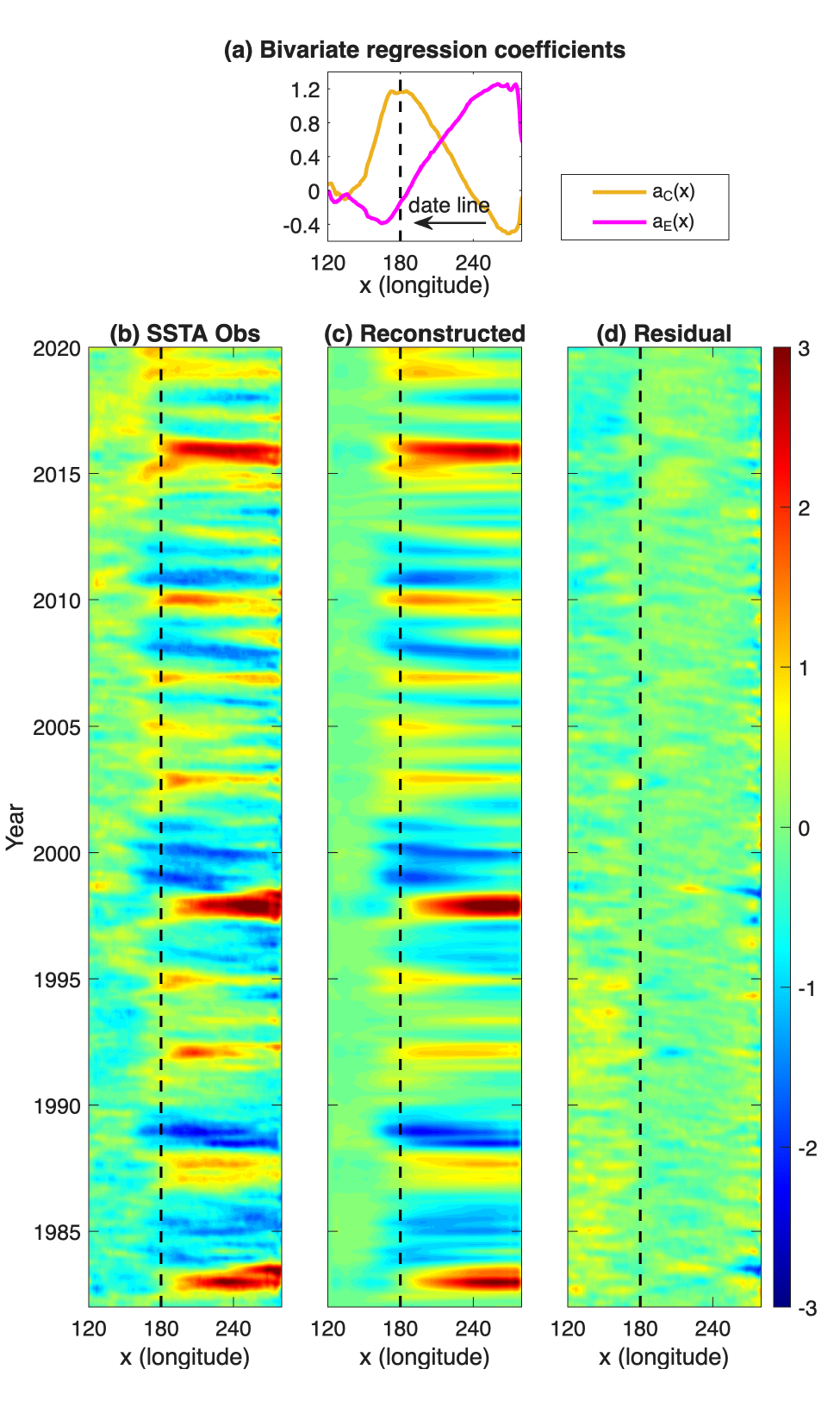}
\caption{Bivariate regression analysis. Panel (a): Regression coefficients $a_E(x)$ and $a_C(x)$ from \eqref{Reconstruction_ENSO}. Panel (b): Observed sea surface temperature (SST) field. Panel (c): Reconstructed SST field based on the bivariate regression using the coefficients from Panel (a). Panel (d): Residual SST, representing the difference between the observed SST in Panel (b) and the reconstructed SST in Panel (c).}\label{fig:SI_Bivariate_Coefficients}
\end{figure}

For each longitude $x^*$, the two scalar regression coefficients, $a_E(x^*)$ and $a_C(x^*)$, are computed using \eqref{Reconstruction_ENSO} based on observational SST anomaly data from 1982 to 2020. Repeating this process across all longitude grids yields the spatially dependent functions $a_E(x)$ and $a_C(x)$. These functions are displayed in Panel (a) of Figure \ref{fig:SI_Bivariate_Coefficients}. The regression coefficient function $a_C(x)$ ($a_E(x)$) peaks in the CP (EP) region, as the SST at those longitudes is closely correlated with $T_C$ ($T_E$).

Given that the spatiotemporal reconstruction based on this bivariate regression is applied to study the coupled ENSO-MJO phenomena, it is crucial to validate the accuracy of the reconstructed field. Panels (b) and (c) of Figure \ref{fig:SI_Bivariate_Coefficients} compare the actual SST field with the reconstructed one, while Panel (d) presents their difference. Overall, the reconstructed field almost perfectly captures the exact SST patterns, especially in the EP and CP regions, with only minor biases in the WP and along the eastern boundary.

\subsection{Reconstruction of the MJO spatiotemporal patterns using Fourier summation}
The reconstruction of the intraseasonal atmospheric variables, which can further describe the spatiotemporal patterns of the MJO, is given by the Fourier summation. For instance, the reconstruction of the moisture variable $Q$ in \eqref{Intraseasonal_atmosphere_model} in physical space is expressed as:
\begin{equation}\label{Reconstruction_MJO}
Q(x,t) = \sum_{k=\pm1,\pm2,\pm3}\hat{Q}_k(t)e^{ikx}.
\end{equation}

\section{Model parameters}\label{sec:params}

The values and definitions of the constant parameters in the conceptual model are listed in Table \ref{params}. We used the values presented in \cite{chen2022multiscale} and \cite{thual2014stochastic} as a starting point for manual tuning. Parameters introduced in the coupling were also fitted manually, starting from values based on physical intuition. Note that the model uses non-dimensional variables, and thus the parameters are unit-less. The dimensional units of each variable are listed in Table \ref{dims}. In addition to those presented in Table \ref{params}, the model uses functions of time and state variables for some parameters. The wind coefficients are important variable parameters. Following the approach of \cite{chen2022multiscale}, the wind forcing is distributed to each state variable of the interannual ocean model in proportion to the components of a coupling vector $e_\beta$, in the order $u, h_W, T_C,$ and $T_E$. In \cite{chen2022multiscale}, these components are obtained directly from the eigenvector of the oscillating ENSO mode of the deterministic system, so that the forcing is imposed along the characteristic direction of the oscillation. Here we use this eigenvector structure as a starting point and adjust the components by manual tuning to reproduce the observed statistics, giving

\begin{equation}\label{eig_betas}
    e_{\beta} = \begin{bmatrix}
        -0.22\\
        -0.53\\
        0.36\\
        1.0
    \end{bmatrix}
\end{equation}

The signs of the components are consistent with the structure of the oscillation, as variables with like signs are positively correlated, while those of opposite sign are negatively correlated. The coupling vector $e_\beta$ gives the following coupling coefficients.

\begin{subequations}
    \begin{align}
        \label{beta_E} \beta_E(I) &= 1.6 - 0.16\,I,\\
        \label{beta_C} \beta_C(I) &= 0.36\, \beta_E(I),\\
        \label{beta_h} \beta_h(I) &= -0.53\, \beta_E(I),\\
        \label{beta_u} \beta_u(I) &= -0.22\, \beta_E(I).
    \end{align}
\end{subequations}

Such functions incorporate decreased sensitivity to changes in wind during periods of strong Walker Circulation, making the regimes associated with the values of $I$ more realistic. Specifically, when the easterly trade winds are strong, the intraseasonal wind $\tau$ is less influential on the interannual ocean model dynamics. This decreases the frequency and strength of EP events. Since the wind plays a more minor role in CP events, and the zonal advection term $\rho I U$ simultaneously strengthens as the wind forcing weakens, CP events increase when background Walker Circulation is strong. In contrast, when $I$ is small, $\tau$ has more influence on the ENSO dynamics, encouraging warming primarily in the EP. The zonal advection in the CP will be weak, but warm events may still occur at lesser relative frequency to EP events as determined by the coupling coefficient $\beta_C$.

The state dependent noise coefficient $\sigma_{I}(I)$ in \eqref{Decadal_variability} can be derived using the Fokker-Planck equation evaluated at the stationary solution of the PDF for I, $p(I)$ \cite{averina1988numerical}. The resulting noise coefficient is

\begin{equation}\label{sigmaI}
    \sigma_I(I) = Re\left( \sqrt{-8 \lambda \int_{0}^I \lambda(y - \overline{I} ) dy}\right).
\end{equation}

\begin{center}
\begin{longtable}{|l|l|l|}
\caption{Definitions and values for the constant parameters in the coupled model.}
\label{params}\\
\hline
Parameter & Definition & Value\\ \hline
\endfirsthead
\hline
Parameter & Definition & Value\\ \hline
\endhead
\hline
\endfoot
\hline
\endlastfoot
$r$ & Collective ocean adjustment rate & 0.0625\\
$r_h$ & Thermocline damping rate & 0.094\\
$\alpha_1$ & Scaling factor to reflect the thermocline feedback on $T_C$ & 0.18\\
$\alpha_2$ & Scaling factor to reflect the thermocline feedback on $T_E$ & 0.15\\
$b_0$ & Upper bound estimation of the thermocline tilt & 2.5\\
$\mu$ & Relative coupling coefficient & 0.5\\
$\gamma$ & Thermocline feedback strength & 0.465\\
$\gamma_C$ & Thermocline feedback strength on $T_C$ & 0.372\\
$\gamma_E$ & Thermocline feedback strength on $T_E$ & 0.256\\
$\delta_{CE}$ & Rescaling factor for the $T_C$ contribution to $T_E$ & 0.10\\
$\rho$ & Coupling coefficient for zonal advection and decadal variability & 0.025\\
$\sigma_C$ & Additive noise amplitude on $T_C$ & 0.05\\
$\sigma_E$ & Additive noise amplitude on $T_E$ & 0.005\\
$C_h$ & Correction term for the $h_W$ mean state & 0.007\\
$C_E$ & Correction term for the $T_E$ mean state & $-0.01$\\
$\lambda$ & Decadal variable damping rate & 0.0333\\
$\overline{I}$ & Decadal variable mean state & 2\\
$d_k$ & Atmospheric damping rate & 4.2\\
$\overline{H}$ & Heating/drying rate & 38.8235\\
$\overline{Q}$ & Background vertical moisture gradient & 0.9\\
$\Gamma$ & Constant of proportionality & 292.9412\\
$\lambda_A$ & Convective activity damping rate & 2\\
$\nu$ & Convective noise scaling factor & 0.25\\
$\tilde{\sigma}_Q$ & Maximum moisture noise amplitude & 2\\
$c_q$ & Latent heat sensitivity & 0.5\\
$\sigma_Q^{\min}$ & Minimum moisture noise amplitude & 0.0067\\
$\alpha_q$ & Latent heat sensitivity to SSTA in the CP & 0.9\\
\end{longtable}
\end{center}

\begin{table}[h!]
    \centering
    \begin{tabular}{|c|c|c|c|c|c|}
    \hline
        Variable & Units&Variable & Units&Variable & Units \\ \hline
         $[U]$& 1.5 m/s  & $[h_W]$ & 150 m & $[T]$ & 7.5 $^\circ$C\\
         $[\tau]$ & 50 m/s & $[Q]$ & 15 K & $[A]$ & 151 $\text{K}^{-1}$ \\
         $[\theta]$ & 15 K & $[t]$ & 2 months & $[x]$ & 15000 km \\
         \hline
    \end{tabular}
    \caption{Dimensional units of  the model variables.}
    \label{dims}
\end{table}

\section{Additional model analysis}\label{sec:additional_analysis}

\subsection{The ENSO-MJO spatiotemporal patterns in observational data}\label{sec:obs_hov}

\begin{figure}[h]
\hspace*{-0.8cm}\includegraphics[width=1.05\textwidth]{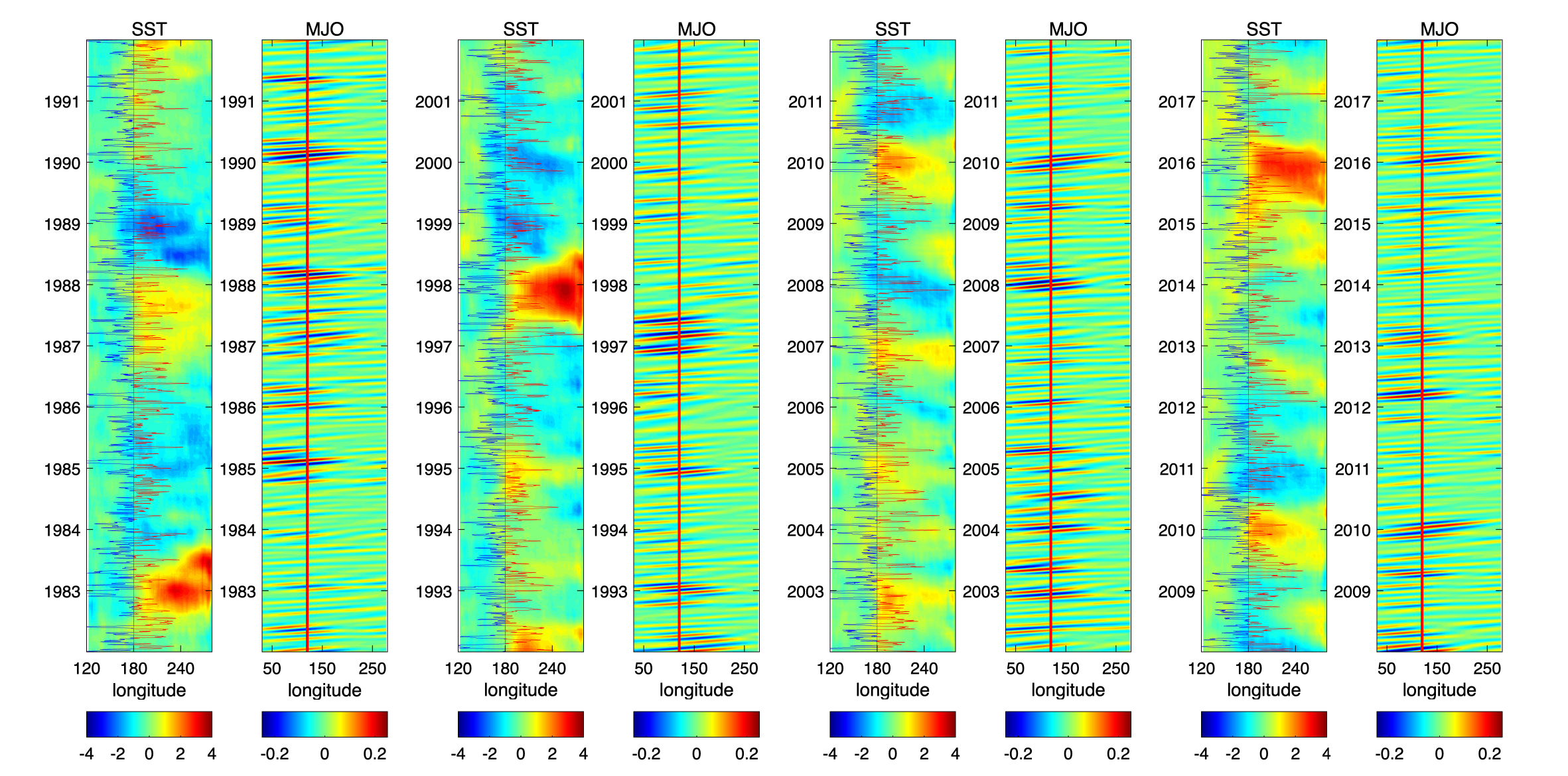}
\caption{Hovmoller diagrams depicting the evolution of ENSO and MJO patterns during the observation period from 1982 to 2018. The format is similar to Figure \ref{fig:Hovmoller_ENSO_MJO}.  }\label{fig:Hov_SST_MJO_obs_new}
\end{figure}

Figure \ref{fig:Hov_SST_MJO_obs_new} presents Hovmoller diagrams of ENSO and MJO patterns during the observational period from 1982 to 2018. Similar to Figure \ref{fig:Hovmoller_ENSO_MJO}, the SST spans the equatorial Pacific, while the MJO also extends into the Indian Ocean. The red vertical line in the MJO panels marks the boundary of the Western Pacific (WP) at 120$^\circ$E. In the SST panels, the averaged atmospheric wind over the WP is overlaid on the SST, with red and blue indicating westerly and easterly intraseasonal wind anomalies, respectively. The black curve represents the interannual wind. This figure is used to qualitatively validate the coupled relationship between ENSO and MJO as illustrated in the stochastic conceptual model shown in Figure \ref{fig:Hovmoller_ENSO_MJO}.

\subsection{Statistics of the other model variables}

\begin{figure}[h]
\hspace*{-0.5cm}\includegraphics[width=1.\textwidth]{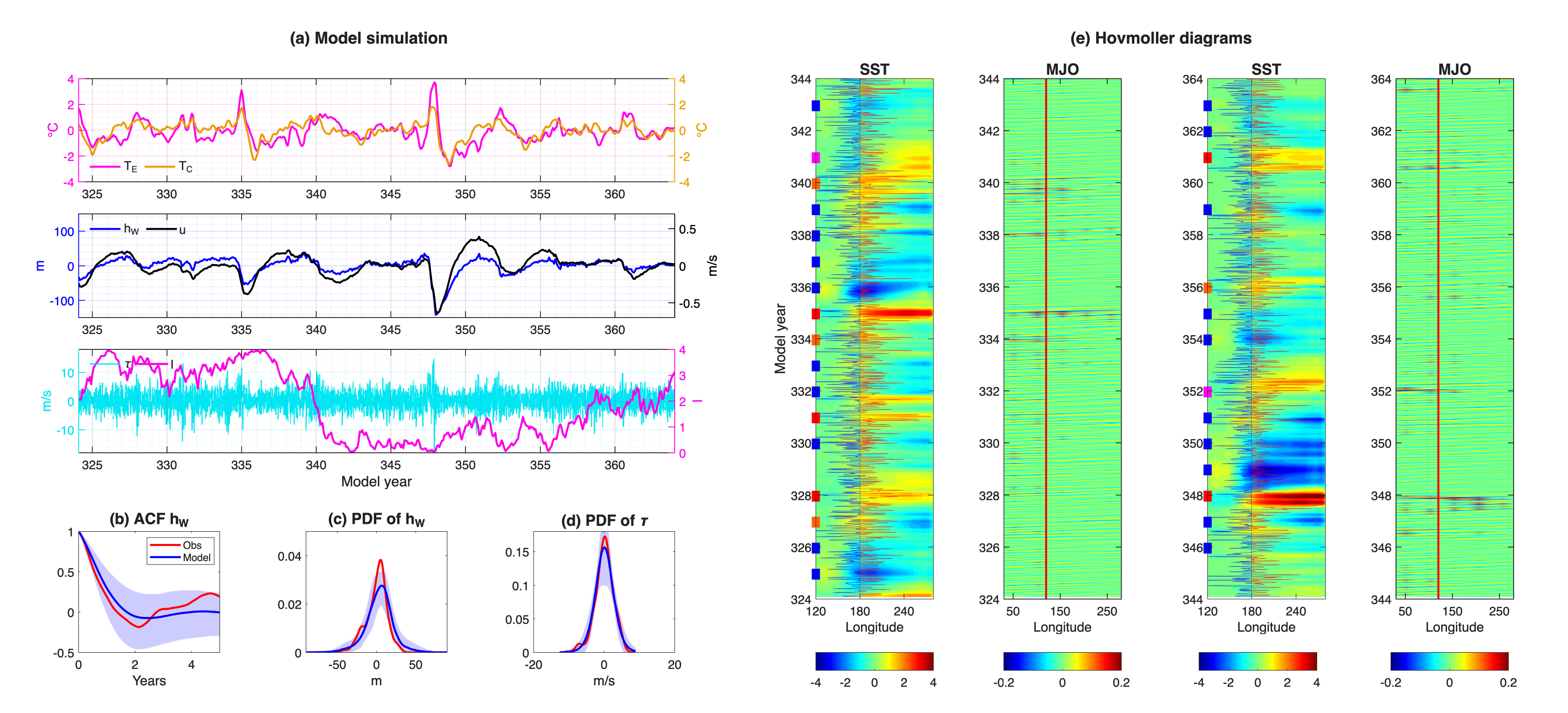}
\caption{Additional model simulation results. Panel (a): Time series of model variables: sea surface temperatures in the eastern and central Pacific ($T_E$ and $T_C$), thermocline depth in the western Pacific ($h_W$), ocean zonal current in the central Pacific ($u$), total wind related to MJO in the western Pacific ($\tau$), and decadal variability ($I$). Panels (b)-(c): ACF and PDF of $h_W$, comparing model results with observations. Panel (d): PDF of $\tau$. Panel (e): Hovmoller diagram of SST ($^\circ$C) and MJO (unit-less) over the same period shown in Panel (a).}
\label{fig:SI_ENSO_Time_Series}
\end{figure}

Figure \ref{fig:SI_ENSO_Time_Series} presents additional model results. While the main text focuses on the two SST variables, this figure includes time series and statistics for other variables in the stochastic conceptual model. The thermocline in the WP, $h_W$, exhibits statistics comparable to observations. The intraseasonal wind statistics closely match the observations, capturing the zero mean and symmetry over time. Additionally, in both observations and the model simulation, the intraseasonal wind on average favors WWBs leading up to EP El Ni\~no events. Accompanying the time series, Hovmoller diagrams of SST and MJO during the same period are also shown. Notably, the decadal variability $I$ modulates the predominant type of El Ni\~no event: during intervals when $I$ is large, the zonal ocean advection is stronger (see \eqref{Interannual_ocean_model_T_C}) and CP El Ni\~no events are favored, whereas when $I$ is small, several strong EP El Ni\~no events occur, along with more active MJO events. This modulation can be seen by comparing the decadal variable $I$ in Panel (a) with the SST Hovmoller diagram in Panel (e). 

\bibliography{references}

\end{document}